\documentclass[10pt,journal]{IEEEtran}
\usepackage{cite}
\usepackage{url}
\usepackage{ragged2e}
\usepackage{footnote}
\makesavenoteenv{tabular}
\makesavenoteenv{table}
\usepackage{epsfig}
\usepackage{graphicx}
\usepackage{amsmath,amssymb} 
\usepackage{algorithm}
\usepackage{algorithmicx}
\usepackage{algpseudocode}
\usepackage{graphics}
\usepackage{threeparttable}
\usepackage{color}
\usepackage[normalem]{ulem}
\usepackage{multirow}
\usepackage{float}
\usepackage{amsfonts}
\usepackage{bm}
\usepackage{array}
\usepackage{pifont}
\usepackage{diagbox}
\usepackage{rotating}
\usepackage{booktabs}
\usepackage{overpic}
\usepackage{textcomp}
\usepackage{contour}

\usepackage{enumitem}
\usepackage{colortbl}
\usepackage[american]{babel}
\usepackage{microtype}
\usepackage{bbding}

\usepackage[pagebackref=false,breaklinks=true,colorlinks, bookmarks=false]{hyperref}

\usepackage{colortbl}
\usepackage{cleveref}
\crefformat{section}{\S#2#1#3} 
\crefformat{subsection}{\S#2#1#3}
\crefformat{subsubsection}{\S#2#1#3}

\usepackage{silence}
\def\ie{\emph{i.e.}}
\def\eg{\emph{e.g.}}
\def\etc{\emph{etc}}
\def\etal{{\em et al.~}}

\newcommand{\myPara}[1]{\vspace{10pt}\noindent\textbf{#1}}

\newcommand{\figref}[1]{Fig.~\ref{#1}}
\newcommand{\tabref}[1]{Table~\ref{#1}}

\newcommand{\secref}[1]{$\S$\ref{#1}}

\def\TotalNumberNeg{75,541~}
\def\PatientAnnoNumberPos{64,771~}
\def\PixelAnnoNumberPos{3,855~}

\def\ourmodel{\textit{JCS}} 
\def\ourdataset{\textit{COVID-CS}}
\graphicspath{{./Imgs/}{../../figure/}{./Imgs/results/}{./Imgs/authors/}}
\begin{document}

\ifdefined \GramaCheck
  \newcommand{\CheckRmv}[1]{}
  \renewcommand{\eqref}[1]{Equation 1}
\else
  \newcommand{\CheckRmv}[1]{#1}
  \renewcommand{\eqref}[1]{Equation~(\ref{#1})}
\fi

\title{JCS: An Explainable COVID-19 Diagnosis System by Joint Classification and Segmentation}

\author{Yu-Huan Wu,
        Shang-Hua Gao,
        Jie Mei,
        Jun Xu,
        Deng-Ping Fan,
        Rong-Guo Zhang,
        and~Ming-Ming~Cheng 
\thanks{Manuscript received April 16, 2020; revised August 11, 2020 and December 14, 2020; accepted February 8, 2021. Date of publication February 18, 2021. 
This work was supported in part by the Major Project for New Generation of AI Grant (No. 2018AAA0100400), NSFC (61922046, 62002176), 
and Tianjin Natural Science Foundation (18ZXZNGX00110). 
(Corresponding author: M.-M. Cheng)}
\thanks{Y.-H.~Wu is with the TKLNDST, College of Computer Science, Nankai University, and the InferVision. (E-mail: wuyuhuan@mail.nankai.edu.cn)
}
\thanks{S.-H. Gao, J. Mei, D.-P. Fan, and M.-M.~Cheng are with the TKLNDST, College of Computer Science, Nankai University. (E-mail: shgao@mail.nankai.edu.cn, meijie0507@gmail.com, dengpfan@gmail.com, cmm@nankai.edu.cn)
}
\thanks{J. Xu is with the School of Statistics and Data Science, Nankai University. (E-mail: nankaimathxujun@gmail.com)
}
\thanks{R.-G. Zhang is with the InferVision. (E-mail: zrongguo@infervision.com)}
}

\markboth{IEEE Transactions on Image Processing, VOL. 30, FEB 2021}
{Wu \MakeLowercase{\textit{et al.}}: JCS: An Explainable COVID-19 Diagnosis System by Joint Classification and Segmentation}

\maketitle

\begin{abstract}
   \justifying
   Recently, the coronavirus disease 2019 (COVID-19) has caused a pandemic disease in over 200 countries, influencing billions of humans.
   To control the infection, identifying and separating the infected people is the most crucial step.
   The main diagnostic tool is the Reverse Transcription Polymerase Chain Reaction (RT-PCR) test. 
   Still, the sensitivity of the RT-PCR test is not high enough
   to effectively prevent the pandemic.
   The chest CT scan test provides a valuable complementary tool to the RT-PCR test,
   and it can identify the patients in the early-stage with high sensitivity.
   However, the chest CT scan test is usually time-consuming,
   requiring about 21.5 minutes per case.
   This paper develops a novel Joint Classification and Segmentation (\ourmodel) system 
   to perform real-time and explainable COVID-19 chest CT diagnosis.
   To train our \ourmodel~system, 
   we construct a large scale COVID-19 Classification and Segmentation (\ourdataset) dataset,
   with 144,167 chest CT images of 400 COVID-19 patients and 350 uninfected cases. 
   \PixelAnnoNumberPos chest CT images of 200 patients are annotated with fine-grained pixel-level labels of opacifications,
   which are increased attenuation of the lung parenchyma. 
   We also have annotated lesion counts, opacification areas, and locations and thus benefit various diagnosis aspects.
   Extensive experiments demonstrate that the proposed \ourmodel~diagnosis system is very efficient for COVID-19 classification and segmentation.
   It obtains an average sensitivity of 95.0\% and a specificity of 93.0\% on the classification test set, 
   and 78.5\% Dice score on the segmentation test set of our \ourdataset~dataset.
   The COVID-CS dataset and code are available at \url{https://github.com/yuhuan-wu/JCS}.
\end{abstract}

\section{Introduction}
\label{sec:introduction}
\IEEEPARstart{C}{oronavirus} disease 2019, or COVID-19, is an epidemic disease caused by the Severe Acute Respiratory Syndrome Coronavirus 2 (SARS-CoV-2).
It outbreaks around the world in a short period 
and has caused 1,914,916  confirmed cases and 123,010 confirmed deaths as of April 15th, 2020.
COVID-19 pushes the health systems of over 200 countries to the 
brink of collapse due to the lack of medical supplies and staff 
and thus has been declared as a pandemic by the World Health Organization~\cite{who2020}.
The current main diagnostic tool for COVID-19 is via 
the Reverse Transcription Polymerase Chain Reaction (RT-PCR) test~\cite{ChineseStandard}.
However, the RT-PCR test is not accurate enough to well prevent the pandemic~\cite{tao2020correlation,fang2020sensitivity}.
So the false-negative cases of RT-PCR tests are a potential threat to public wellness, and missing any COVID-19 cases will probably cause secondary infections of large areas.
%
%

\CheckRmv{
\begin{figure}[t!]
	\centering
    \small
	\begin{overpic}[width=\columnwidth]{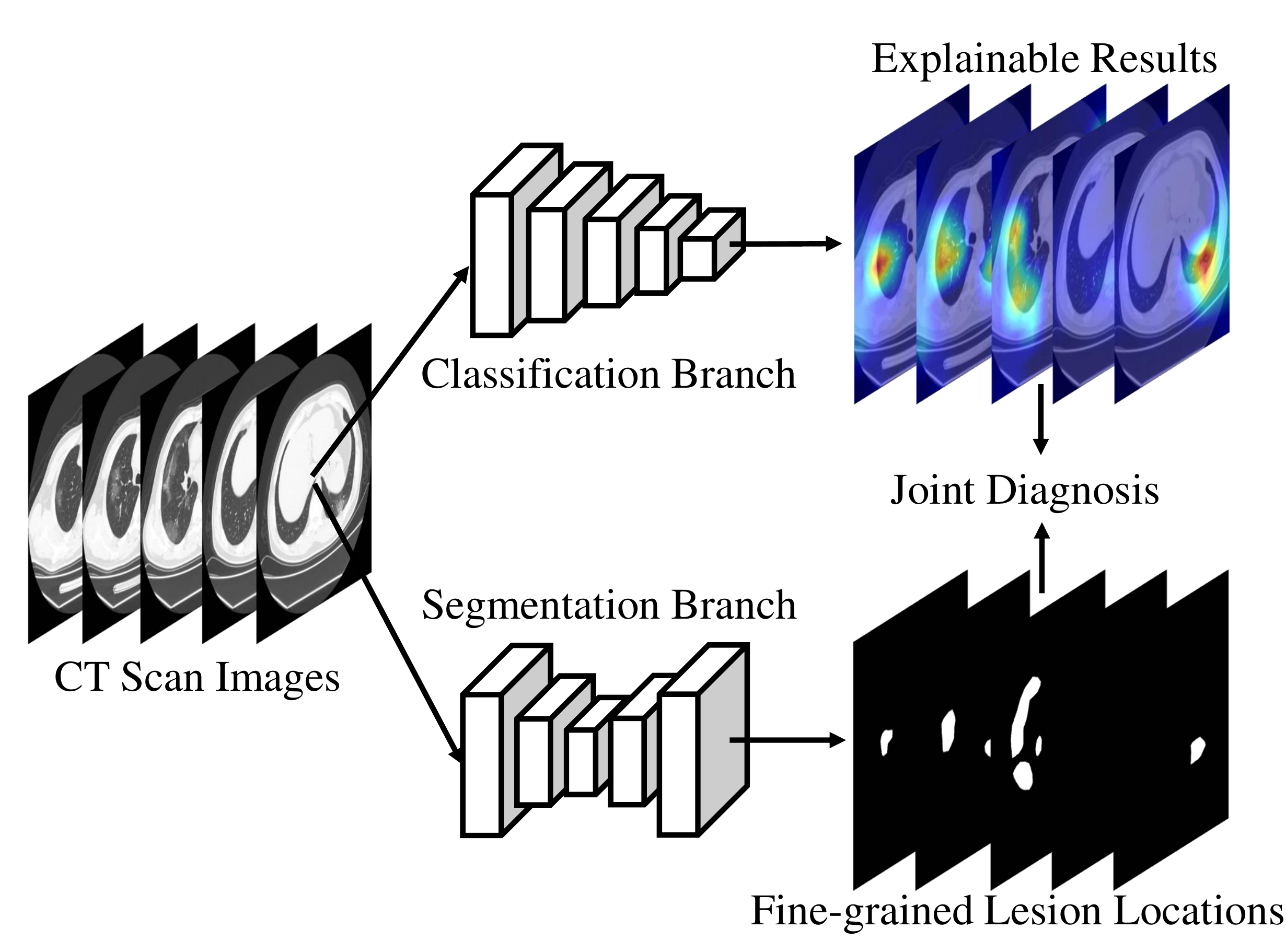}
    \end{overpic}
	\caption{\textbf{Illustration of our JCS diagnosis system for COVID-19}.
	Our JCS system will perform the segmentation diagnosis only if the classification branch reports positive COVID-19 predictions.} 
    \label{fig:demo}
\end{figure}
}

To hinder the terrific infection of COVID-19, medical radiology imaging is employed as a complementary tool for the RT-PCR test~\cite{wang2020combination}.
This is based on the fact that the clinical signs of chest X-rays for most COVID-19 patients indicate lung infection~\cite{zhang2020covid19}.
The works of~\cite{tao2020correlation, fang2020sensitivity} show that CT scan tests are with high sensitivity.
Besides, a CT scan test is necessary for monitoring the severity of the illness~\cite{inui2020chest}.
However, the diagnosis duration is the major limitation of CT scan tests:
even experienced radiologists need about 21.5 minutes~\cite{huang2020battle}
to analyze the test results of each case.
The experienced radiologists are severely lack during the pandemic outbreak,
posting difficulties identifying potentially infected patients in time.
Thus, automatic diagnosis systems are highly desired.

Thanks to the powerful discriminative ability of deep convolutional neural networks (CNNs), 
artificial intelligence (AI) technologies are reforming the medical imaging community.
Deep CNNs are usually trained on large scale datasets to demonstrate their capability.
However, most of the existing CT scan datasets for COVID-19~\cite{cohen2020covid,zhao2020COVID-CT-Dataset,PLXR,CTSeg} could not meet this demand, as they contain at most hundreds of CT images from tens of cases.
Besides, most of the current COVID-19 datasets only provide the patient-level labels (\ie, class labels) 
indicating whether the person is infected and lacks fine-grained pixel-level annotations.
Thus, CNN models trained with these datasets usually neglect the valuable information for explaining the final predictions.
Despite several CT scan diagnosis systems~\cite{AIDistinguish2020,Chung2020, Bernheim2020,wang2020,shi2020,fang2020sensitivity} have been established for testing the suspected COVID-19 cases, most of them suffer from two drawbacks:
1) they are trained on small scale datasets and thus not robust enough for versatile COVID-19 infections;
2) they perform classification based on the black box CNNs while lacking the explainable transparency to assist the doctors during the medical diagnosis.

To alleviate the  drawbacks mentioned above, in this work, we
1) construct a large scale \ourdataset~dataset with both patient-level and pixel-level annotations and 
2) propose a \textit{J}oint \textit{C}lassification and \textit{S}egmentation (\ourmodel) based diagnosis system to provide explainable diagnosis results for medical staffs fighting with COVID-19.
Specifically, we utilize the collected \ourdataset~dataset that contains thousands of CT images from hundreds of COVID-19 cases to train our \ourmodel~system for better diagnosis performance.
As illustrated in Figure~\ref{fig:demo}, 
our \ourmodel~diagnosis system first 
identifies the suspected COVID-19 patients by a classification branch 
and provides diagnosis explanations via activation mapping techniques \cite{Selvaraju_2017_ICCV}.
Our system is then feasible to discover the locations and areas 
of the COVID-19 infection in lung radiography via fine-grained image segmentation techniques.
With the explainable classification results and corresponding fine-grained lesion segmentation, our \ourmodel~system largely simplifies and accelerates the diagnosis process for radiologists or other medical experts.

As shown in \tabref{tab:DiagnosisTime}, our \ourmodel~system needs only 22.0 seconds for each infected case or 1 second for each uninfected case,
much faster than the RT-PCR tests and CT scan analysis by experienced radiologists.
With the assistance of our \ourmodel~system, experienced radiologists only cost 54.4 
(32.4 for radiologists and 22.0 for \ourmodel) seconds 
for each infected case or 1.0 second for each uninfected case, keeping the same high specificity and sensitivity. 
Hence, the speed and effectiveness of assistance have shown the superiority of our \ourmodel~system.

In summary, our contributions are mainly three-fold:
\begin{itemize}
    \item \textbf{We construct a new large scale COVID-19 dataset}, called \ourdataset, 
    which contains \PixelAnnoNumberPos fine-grained pixel-level labeled CT images from 200 COVID-19 patients, 
    \PatientAnnoNumberPos patient-level annotated CT images from 200 other COVID-19 patients, 
    and \TotalNumberNeg CT images of 350 uninfected cases.
    \item \textbf{We develop a novel COVID-19 diagnosis system} to perform explainable Joint  Classification and accurate lesion Segmentation (\ourmodel), showing clear superiority over previous systems.
\item On our \ourdataset~dataset, \textbf{our \ourmodel~system achieves 95.0\% sensitivity and 93.0\% specificity on COVID-19 classification, and 78.5\% Dice score on segmentation}, surpassing previous state-of-the-art segmentation methods.
\end{itemize}

The remaining paper is organized as follows.
In \S\ref{sec:related}, we briefly summarize the related works.
In \S\ref{sec:system}, we introduce the developed diagnosis system for recognizing and analyzing the COVID-19 cases.
In \S\ref{sec:dataset}, we present our \ourdataset~dataset with our labeling procedures in detail.
Extensive experiments are conducted in \S\ref{sec:experiment} to evaluate the performance of our system on COVID-19 recognition, with in-depth analysis. 
\S\ref{sec:conclusion} concludes this work.

\CheckRmv{
\begin{table}[t!]
  \centering
  \small
  \setlength\tabcolsep{3pt}
  \caption{\small
  Summary of different datasets (updated on 2020/4/10). 
  }
  \vspace{-1mm}
  \label{tab:DatasetSummary}
  \begin{tabular}{r||c|c|c|c|r} \hline
   Dataset~~~~ & Date & Link & Type & \#Images & \#Cases \\
  \hline
  \hline
  PLXR~\cite{PLXR} & 2020/03/23 & \href{https://www.kaggle.com/nabeelsajid917/covid-19-x-ray-10000-images}{Link} & X-rays & 98 & 70 \\
  8023Dataset~\cite{cohen2020covid} & 2020/03/25 &  \href{https://github.com/ieee8023/covid-chestxray-dataset/tree/3f07c70b7727b4695b4c89a499ae743d16c3caa7}{Link} & X-rays & 229$^*$ & - \\
  CTSeg~\cite{CTSeg} & 2020/03/28 &
  \href{http://medicalsegmentation.com/covid19}{Link} & CT & 110 & 60 \\
  COVID-CT~\cite{zhao2020COVID-CT-Dataset} & 2020/03/30 & \href{https://github.com/UCSD-AI4H/COVID-CT/tree/7f65bf2c99b0909d919d43a5f53be70d4e77440b}{Link}  & CT & 746{$^*$} & - \\
  \hline
  \textbf{\ourdataset~(Ours)} & 2020/04/12 & - & CT &
  \textbf{$>$144K$^\dagger$} & \textbf{750} \\
  \hline
  \multicolumn{6}{p{246pt}}{$^*$: The number is reported from the authors' GitHub repository.}\\
  \multicolumn{6}{p{246pt}}{$^\dagger$: Among our dataset, \PixelAnnoNumberPos images of 200 positive cases are pixel-level annotated,
  \PatientAnnoNumberPos images of the other 200 positive cases are patient-level annotated, and the rest \TotalNumberNeg images are from the 350 negative cases.}
  \end{tabular}
\end{table}
}

\CheckRmv{
\begin{table}[t!]
  \centering
  \normalsize
  \setlength\tabcolsep{3pt}
  \caption{
  Average time of COVID-19 diagnosis by different methods.
  ``CT R.'' indicates CT radiologist and ``CT R. + \ourmodel'' is CT radiologist diagnosis with the assistance of \ourmodel.
    }
 \vspace{-1mm}
  \label{tab:DiagnosisTime}
  \begin{tabular}{r||c|c|c|c} \hline
  Method & RT-PCR & CT R. & CT R. + \ourmodel & \ourmodel
  \\
  \hline
  Time & $\sim$4h~\cite{won2020development} & 21.5min~\cite{huang2020battle} & 1s$^\dagger$/54.4s & 1s$^\dagger$/22.0s 
  \\
  \hline
  \multicolumn{5}{p{246pt}}{$\dagger$: diagnose uninfected cases.}
  \end{tabular}
\end{table}
}

\section{Related Works}
\label{sec:related}
\subsection{Existing Accessible COVID-19 Datasets}
As of April 15th, 1,914,916 people are infected by COVID-19.
But their CT scans are usually private and could not be publicly accessed.
To facilitate the development of diagnostic systems,
several COVID-19 related datasets are publicly released by researchers around the world.
A summary of these datasets is provided in Table~\ref{tab:DatasetSummary}.

One X-ray dataset from Cohen \etal\cite{cohen2020covid} contains overall 122 frontal view X-rays, including 100 images of COVID-19 cases, 11 SARS images, and 11 other pneumonia images.
The COVID-CT dataset from~\cite{zhao2020COVID-CT-Dataset} has 746 CT scan images, 
349 images from COVID-19 patients and 397 from non-COVID-19 cases.
All the images in these datasets are collected from public websites and/or COVID-19 related papers on medRxiv, bioRxiv, and journals, \etc.
CTs containing COVID-19 abnormalities are selected by reading the figure captions in the papers. 
Some other resources of the COVID-19 dataset are PLXR~\cite{PLXR} and CTSeg~\cite{CTSeg}, which contains 98 and 110 CT scan images cases, respectively.
These datasets are on a small scale and lack diversity
since they often contain less than hundreds of images from tens of cases.
To fully exploit the power of deep CNNs, it is essential to construct a
large scale dataset to train deep CNNs in accurate and robust COVID-19 systems.

\subsection{Manual COVID-19 Diagnosis}

The most crucial step of preventing the spread of the COVID-19 is 
immediately identifying every patient from normal people.
Missing any patient will probably cause secondary COVID-19 infections 
in large areas.
Currently, the main manual diagnostic tool is the RT-PCR test~\cite{udugama2020diagnosing}.
However, the sensitivity of RT-PCR test is not high enough to effectively 
prevent the pandemic~\cite{tao2020correlation,fang2020sensitivity}.
As widely available in many hospitals, CT scan is a complementary tool to the RT-PCR test. However, some special cases with the RT-PCR test confirmed positive have normal CTs~\cite{yang2020patients,hu2020clinical,bernheim2020chest}.
Combining both tests allows maximally to identify potentially infected people, 
as it can identify COVID-19 patients in the early-stage with high 
sensitivity~\cite{tao2020correlation,long2020diagnosis,fang2020sensitivity}.
The CT scan is also necessary for monitoring the severity of 
the illness~\cite{inui2020chest}.
During the pandemic outbreak, experienced medical staff
is severely lacking,
posting difficulties identifying potentially infected patients in time.
Thus, automatic diagnosis systems are highly desired.

\subsection{Automatic COVID-19 Diagnosis Systems}
Most current medical imaging systems are developed for common diseases that exist for many years, \eg, tuberculosis~\cite{TBX11K}.
These developed systems can be directly modified to attenuate the COVID-19 outbreak.
The doctors find that the chest X-rays of COVID-19 patients exhibiting certain abnormalities in the radiography.
Based on ResNet-50~\cite{he2016deep}, COVID-ResNet~\cite{Farooq2020COVIDResNetAD} is proposed to differentiate 
three types of COVID-19 infections from normal pneumonia individuals.
On 1531 chest X-ray images, Zhang \emph{et al.} proposed a deep anomaly detection system for COVID-19 screening, achieving 96.0\% sensitivity and 70.65\% specificity.
Yang \etal\cite{yang2020} proposed a system to evaluate the images of 102 volunteers, with a sensitivity of 83.3\% and specificity of 94.0\%.
The system developed by Li \etal\cite{Li2020} identifies 78 COVID-19 patients with a sensitivity of 82.6\% and a specificity of 100.0\% by using the axial and coronal-view of lung CT severity index (CTSI).
Chung \etal\cite{Chung2020} confirmed via collected from 21 patients that
visual inspection helps identify the COVID-19 cases and predict the severity via the overall lung total severity score (LTSS).
Bernheim \etal\cite{Bernheim2020} analyzed the 121 COVID-19 patients 
and carried on a visual check by the experienced radiologist to categorize them as early, intermediate and late cases.
Wang \etal\cite{wang2020} found that the COVID-19 disease will be severe during 6-11 days from the infection, based on a study on 366 CT scans of 90 patients.
Shi \etal\cite{shi2020} developed an imaging-assisted diagnosis procedure to detect the COVID-19 caused pneumonia.
Fang \etal\cite{fang2020sensitivity} examined 81 patients by a procedure based on the CTSI 
and obtained a sensitivity of 98.0\%, 
in contrast to the sensitivity of 71.0\% by RT-PCR.
Zhou \etal\cite{zhou2020} implemented the examination using the non-contrast CTSI of 62 COVID-19 patients, 
confirming that the CT-assisted evaluation shows better detection accuracy in the progressive stage confirmed to the early-stage.
Despite their success on a small set of samples, 
these COVID-19 diagnosis systems have not been tested by large scale samples. 
They could not provide useful diagnostic evidence during their diagnostic inference.
More works can refer to the reviews of \cite{wynants2020prediction,fan2020inf,qiu2020miniseg}.

As far as we know, 
only two works extract infected regions via pixel-level segmentation.
Rajinikanth~\etal\cite{rajinikanth2020harmonysearch} performed the segmentation
via the watershed transform techniques~\cite{watershed2000} with coarse results and limited accuracy.
Zhou~\etal~\cite{zhou2020unet} developed a U-Net with an attention mechanism and obtained a Dice score of 69.1\% on CTSeg~\cite{CTSeg} dataset, but its training and test split have only 88 and 22 images.
In this work, we propose a diagnosis system 
by integrating learning-based classification and segmentation networks
to provide explainable diagnostic evidence for doctors 
and improve the user-interactive aspects of the diagnosis process.

\subsection{Deep Classification and Segmentation Methods}

Ever since the release of the ImageNet dataset~\cite{imagenet}, 
deep convolutional neural networks (CNNs) have become
the workhorse for image classification tasks with improving performance.
Representative deep classifiers, \eg, AlexNet~\cite{alexnet}, VGGNet~\cite{simonyan2015very}, ResNet~\cite{he2016deep}, DenseNet~\cite{densenet}, and Res2Net~\cite{pami20Res2net}, have been widely employed as the feature extractors for other computer vision tasks,
such as image segmentation~\cite{cheng2016hfs,liu2018deep, zhao2018flic, jiang2021online}, 
visual saliency~\cite{liu2019simple, zhao2019egnet, liu2020light, wu2020edn, wu2020mobilesal}, 
face recognition~\cite{zhao2019regularface}, 
aerial images analysis~\cite{VecRoad_20CVPR}, 
style transfer~\cite{cheng2019structure}, 
feature matching~\cite{bian2019evaluation}, 
crowd counting~\cite{zhang2019nonlinear},
and image restoration~\cite{xu2019nac}, \emph{etc}.
Despite impressive representation ability of these
classifiers, the classification process does not explain clearly the predicted results.

Image segmentation tackles the problem of pixel-level predictions.
Semantic segmentation aims to classify the semantic label for each pixel on a natural image~\cite{wang2019weakly}.
Representative works in this area include FCN~\cite{long2015fully} and DeepLab~\cite{DeepLab2017}. 
Instance segmentation focuses on discriminating each semantic instance with a unique instance label and pixel-level mask in the image~\cite{21PAMI_InsImgDatasetWSIS,wu2021regularized,he2017mask}.
Panoptic segmentation~\cite{sofiiuk2019adaptis} integrates semantic segmentation and instance segmentation, 
and it does semantic segmentation on non-objects (sky, water, grass, etc.) and instance segmentation on objects (cat, dog, bus, etc.).
U-Net~\cite{unet} is a widely employed network for medical image segmentation analysis.
It is further extended to 3D U-Net~\cite{3dunet}, TernausNet~\cite{TernausNet}, and U-Net++~\cite{zhou2019unetplusplus} with promising performance on versatile image segmentation tasks.
In this work, we develop a novel COVID-19 diagnosis system by integrating deep image classification and segmentation techniques.

\section{Our COVID-19 Diagnosis System}
\label{sec:system}

The opacification is the basic CT feature of COVID-19 patients~\cite{xiong2020clinical}, 
and it is defined as the increased attenuation of the lung parenchyma~\cite{leung1993parenchymal}.
Our \ourmodel~system consists of an explainable classification branch to identify the COVID-19 opacifications and a segmentation branch to discover the opacification areas.
The classifier is trained on many images with low-cost patient-level annotations 
and some images with pixel-level annotations for better activation mapping.
And the segmentation branch is trained with accurately annotated
CT images, performing fine-grained lesion segmentation.
By integrating the two models, our \ourmodel~system provides informative diagnosis results for COVID-19.
 
\subsection{Explainable Classification}
Owing to the strong representation ability of CNNs, 
the COVID-19 infections can be predicted through only patient-level supervised training.
To this end, we build a classification branch that consists of the proposed classification model to endow our \ourmodel~diagnosis system 
with the capability of discriminating the COVID-19 patients.

\subsubsection{Diagnosing COVID-19 via Classification} 
Predicting whether the suspected patient is COVID-19 positive or not is 
a binary classification task based on his/her CT scan images.
Since designing the novel classification model is not our focus, 
we build our classifier based on the Res2Net network~\cite{pami20Res2net}.
As a powerful network, Res2Net has a stronger multi-scale representation ability than ResNet~\cite{pami20Res2net}.
%
%
The last layer is modified as a fully-connected layer with two channels 
to output the probability of COVID-19 infection or not.
If the probability of the infected channel is larger than that of the uninfected one, the patient is diagnosed as COVID-19 positive, or vice versa.
For each patient, the CT images are sent to the classification model one by one.
If the number of infected CT images is above a threshold,
the patient is diagnosed as COVID-19 positive.

\subsubsection{Explanation by Activation Mapping} 
As the diagnosis process of CNN classification is in a black box, we employ the activation mapping~\cite{Selvaraju_2017_ICCV} to increase the explainable transparency of our COVID-19 diagnosis system on its predictions.
The last convolutional layer of the classification network is followed by a global average pooling (GAP) layer
and a fully-connected layer.
Through the GAP layer, our classification model down-samples the feature size from $(H, W)$ to $(1, 1)$, and thus lost the spatial representation ability.
Through activation mapping~\cite{Selvaraju_2017_ICCV}, our system finds the response region of the prediction result. 
The hypothesis is that the gradient of regions in features before the GAP layer is consistent with the prediction evidence.
The feature map before the GAP layer contains both high-level semantic and location information.
Each channel corresponds to the activation of different semantic cues.
The activation mapping is obtained through the gradients of the predicted probability of the feature map.
Specifically, given the prediction of COVID-19 branch $y_p$ and the feature map $X$ before GAP,
the weight for the $k$-th channel of $X$ is calculated as:
\begin{equation}
    w_k = \frac{1}{HW} \sum_{i=1}^{H}\sum_{j=1}^{W} \frac{\partial y_p}{\partial X_{i,j}^{k}},
    \label{eqn:gradient}
\end{equation}
where $X_{i,j}^k$ is the value at position $(i,j)$ in the $k$-th channel of feature map $X$.
Larger gradients in Eqn.~(\ref{eqn:gradient}) produce a larger weight of the activation mapping for a certain channel.
The activation mapping for a COVID-19 case is computed as:
\begin{equation}
    AM_p = \sum_{k} ReLU (w_k X^{k}).
\end{equation}
As shown in~\figref{fig:am}, the activation mapping accurately locates the opacification areas of COVID-19 patients, providing explainable results for the prediction of our \ourmodel~system.

\subsubsection{Alleviating Data Bias by Image Mixing}

By utilizing our explainable classification model, our system can be trained only with patient-level annotation.
However, since CT images are from multiple sources, the classifier may be 
trained to overfit unwanted areas (e.g., the area outside the lesion), 
as observed via the activation mapping.
Therefore, we propose to utilize the image \emph{mixing} technique~\cite{zhang2018mixup} and help the classifier focus on the lesion areas of COVID-19 cases.
The CT images from different sources and the corresponding patient-level annotations are mixed during training.
Specifically, for two randomly sampled CT images $x_i$ and $x_j$ ($i\neq j$) and corresponding labels $\hat{y}_i$ and $\hat{y}_j$, the newly mixed sample and the corresponding label are written as:
\begin{eqnarray}
  \begin{aligned}
    x^{m}_{ij} = \lambda x_i + (1-\lambda) x_j, \\
   \hat y^{m}_{ij} = \lambda \hat{y}_i + (1-\lambda) \hat{y}_j,
  \end{aligned}
  \label{eq:global_mix}
\end{eqnarray}
where $\lambda\in[0, 1]$ is a random number generated in Beta distribution, \ie, $\lambda\sim \text{Beta}(\alpha,\alpha)$.
With mixed samples, our classification model is trained to focus more on the decisive lesion areas of COVID-19 cases, rather than the bias in the data source.
Also, the mixing process weakens the confidence of labels, and thus alleviating our system from overfitting.

\subsubsection{Pixel-level Supervision for Activation Mapping}\label{sec:cls_pixel}
Traditional classification models only utilize image labels for training. 
The activation mapping of them may be inaccurate as these models automatically learn the differences of images of different classes.
In our proposed dataset, there are thousands of images with pixel-level annotations for the specific opacification areas, 
and they can be the direct supervision of the activation mapping.
Motivated by the above observations and the work of~\cite{li2019guided}, during the training network, 
we apply a segmentation loss $L_{seg}$ for the activation mapping of the COVID-19 class channel:
\begin{equation}
	L_{seg} = \frac{1}{HW}\|AM_{p,c}^{norm} - S\|_{2},
\end{equation}
where $AM_{p,c}^{norm}$ is the activation mapping of the COVID-19 class channel normalized to $(0,1)$, $S$ is the binary ground truth pixel-level annotation map,  $\|\cdot\|_2$ denotes the 
$\ell_{2}$ norms. $L_{seg}$ will not be computed if images have no ground truth pixel-level annotations.
After applying the segmentation loss $L_{seg}$, \figref{fig:am} shows that the activation mapping significantly improves in locating opacifications.

\subsection{Accurate Segmentation}\label{sec:segmentation}

\CheckRmv{
\begin{figure*}[t]
	\centering
    \small
	\includegraphics[width=\textwidth]{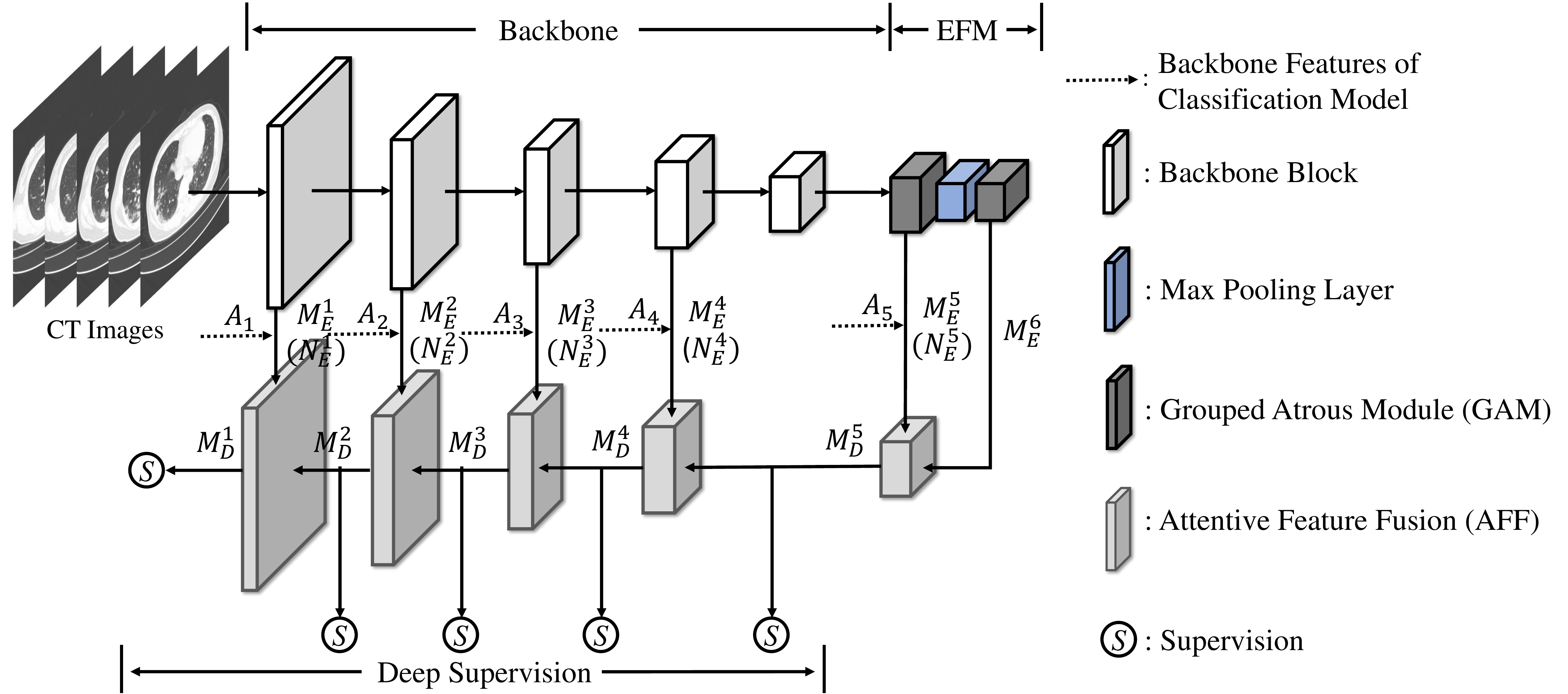}
	\caption{\small \textbf{The architecture of our segmentation branch}.
	EFM indicates the Enhanced Feature Module (\secref{sec:efm}). 
	AFF refers to the Attentive Feature Fusion strategy (\secref{sec:aff}). 
    If not combined with the classification model, $M_E^1 \sim M_E^5$ will be fed into the decoder; 
    otherwise, the combined $N_E^1 \sim N_E^5$ will be fed into the decoder (\figref{fig:Combination}, \secref{sec:combine}).
	We apply deep supervision to train our segmentation branch (\secref{sec:deep_supervision}).
	}
    \label{fig:seg_pipeline}
\end{figure*}
}
\vspace{0.01mm}

Our segmentation branch aims at discovering the exact lesion areas from the CT images of COVID-19 patients.
\figref{fig:seg_pipeline} shows the architecture of our segmentation branch with or without the combination of the segmentation and classification models. 
The details of such a combination are illustrated in \figref{fig:Combination}.

\subsubsection{Encoder-Decoder Architecture}\label{sec:pipeline}

Our segmentation model consists of an encoder and a decoder.

\CheckRmv{
\begin{figure}[t] 
	\centering
    \small
	\begin{overpic}[width=\linewidth]{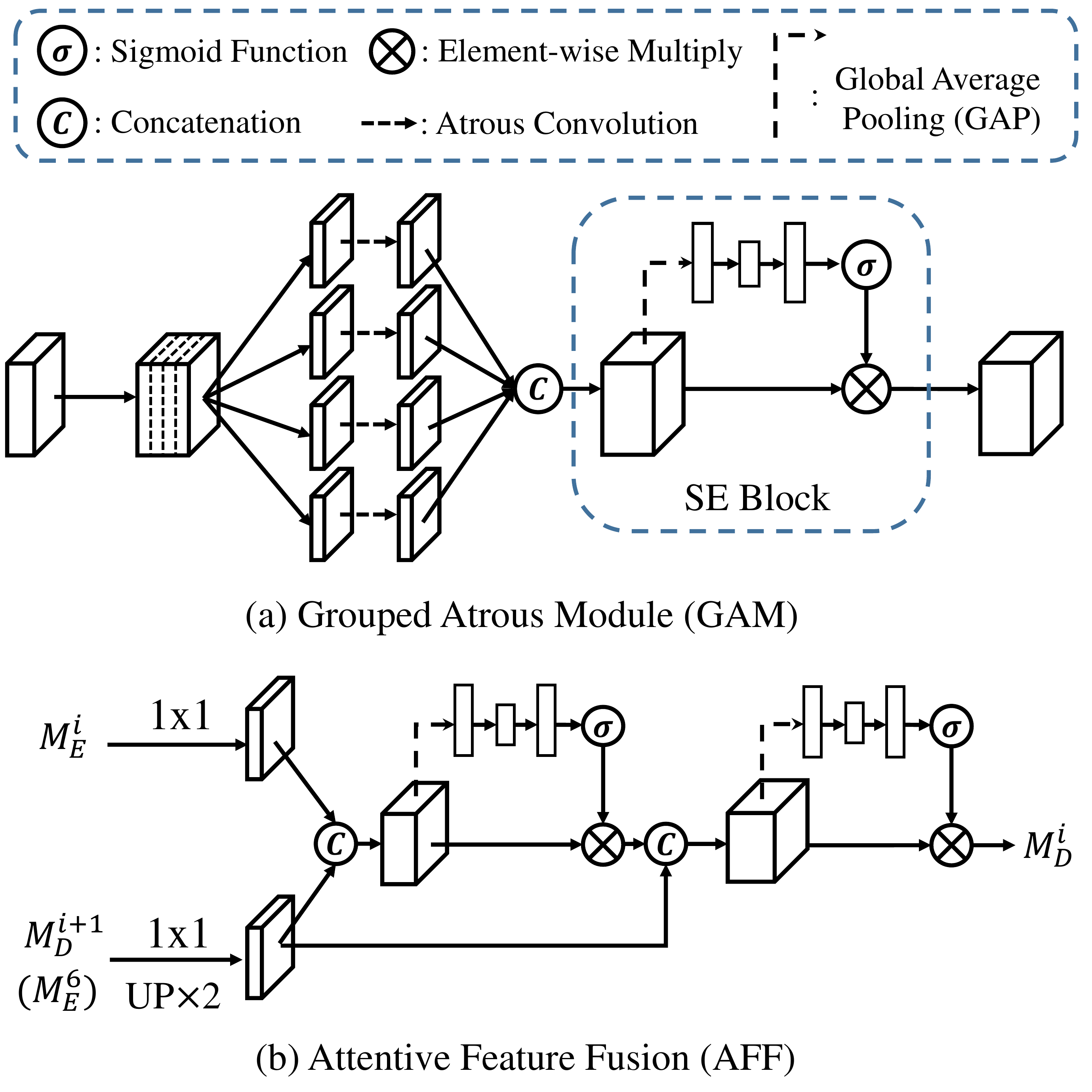}
    \end{overpic}
	\caption{\textbf{Proposed (a) GAM and (b) AFF for the segmentation network}.
	In AFF, $M_D^{i+1}$ will be replaced with $M_E^6$ if $i=5$. 
	Cubes represent three-dimensional feature maps, while rectangles mean feature vectors.}
    \label{fig:TwoModules}
\end{figure}
}

\myPara{Encoder}.
The encoder is based on the VGG-16~\cite{simonyan2015very} backbone, without the last two fully-connected layers.
It has five VGG blocks defined as $\{E_1, E_2, E_3, E_4, E_5\}$, respectively.
The VGG-16 backbone is first fed with the CT images and 
produces multi-scale feature maps from the last layers of the five VGG blocks.
To downsize the input feature map by half, the front of each block (except the first one) is a $max~pooling$ function with a stride of 2.
The feature map produced by the block $E_1$ contains the finest features with the highest resolution, while the feature map by the block $E_5$ is coarsest with the lowest resolution.
To achieve better performance, we propose an Enhanced Feature Module 
(EFM, which will be introduced in \S\ref{sec:efm}) 
for our encoder to improve its representational power.
The EFM module is added after the last layer $conv5\_3$ in the block $E_5$.
It consists of two Grouped Atrous Modules (GAM) to extract stronger feature maps with larger receptive fields.
The GAM module generates an extra smaller feature map, 
half size compared to the coarsest feature map of the VGG-16 backbone.
It also enhances the representational power of the feature map produced by the block $E_5$.
Hence, our encoder produces six levels of feature maps $\{M_E^1, M_E^2, M_E^3, M_E^4, M_E^5, M_E^6\}$, with strides of $\{1, 2, 4, 8, 16, 32\}$, respectively.
As we employ a U-shape encoder-decoder architecture~\cite{ronneberger2015unet}, all these six feature maps will be used in the decoder, as will be introduced later.

\myPara{Decoder}.
Our decoder has five side-outputs with 5 different sizes.
Here, we do not predict the side-output from the coarsest feature map 
with a stride of 32,
and thus no side-output matches the size of the coarsest feature map $M_E^6$.
In our decoder, we propose an Attentive Feature Fusion (AFF, 
which will be introduced in \S\ref{sec:aff}) 
strategy to aggregate the feature maps from different stages and predict the side-output of each stage.
Our AFF emphasizes the significance of the top-level feature map and 
utilizes the attention mechanism to filter useful features from the bottom feature map.
The last output with the same resolution of the CT image input will be chosen as the final prediction.

\subsubsection{Enhanced Feature Module}
\label{sec:efm}
The proposed EFM module is added after the last layer of $E_5$ in the VGG-16 encoder.
It consists of two sequential GAM modules and a $max~pooling$ function between them.
As shown in \figref{fig:TwoModules} (a), 
the first layer of the GAM module is a $1\times 1$ convolution layer to expand the channels of the feature map.
Then the feature map is equally divided into 4 groups.
Unlike the trivial group convolution,
we deploy atrous convolution~\cite{DeepLab2017} with different atrous rates 
to the 4 groups to derive a more abundant feature map with various receptive fields.
Atrous convolution can greatly enlarge the perceptive field of convolutional filters and keep the same computational cost with normal convolution. In 2D cases, atrous convolution with $3\times 3$ kernel size can be simply formulated as below:
\begin{equation}
    q[i,j] = bias + \sum_{k=-1}^{+1}\sum_{l=-1}^{+1}(x[i+k\cdot n,j+l\cdot n] \cdot w[k+1,l+1]) ,
\end{equation}
where n indicates the atrous rate, $w$ is the convolution weight of which the size is $3\times 3$, 
$q$ and $x$ are output and input feature map, respectively, $i$ and $j$ are the feature map location. 
Note that $n=1$ is the special case for normal convolution.
To fully exploit useful features, we adopt the Squeeze-Excitation (SE) block~\cite{hu2018squeeze} in our network, 
that is, using the attention mechanism for re-calibrating channel-wise convolutional feature responses.
More specifically, 
each channel of the input feature map will be multiplied by a channel factor calculated by a SE block.
The SE block consists of two linear layers, followed by a sigmoid function.
The input feature map after global average pooling will be fed into 
this block and we can derive a channel factor ranging $(0,1)$ for each input feature channel.
We set the reduction rate in the SE block as 4, which means 
we set the output number of the first linear layer as the $1/4$ number of the input channels.
To reduce the output channels by half, we add a $1\times 1$ convolution layer after the SE block.

At last, we use a $3\times3$ convolution layer, in which the number of channels equals that of the input feature map, as the transition layer to the next module.

\CheckRmv{
\begin{figure}[t] 
	\centering
    \small
	\begin{overpic}[width=\linewidth]{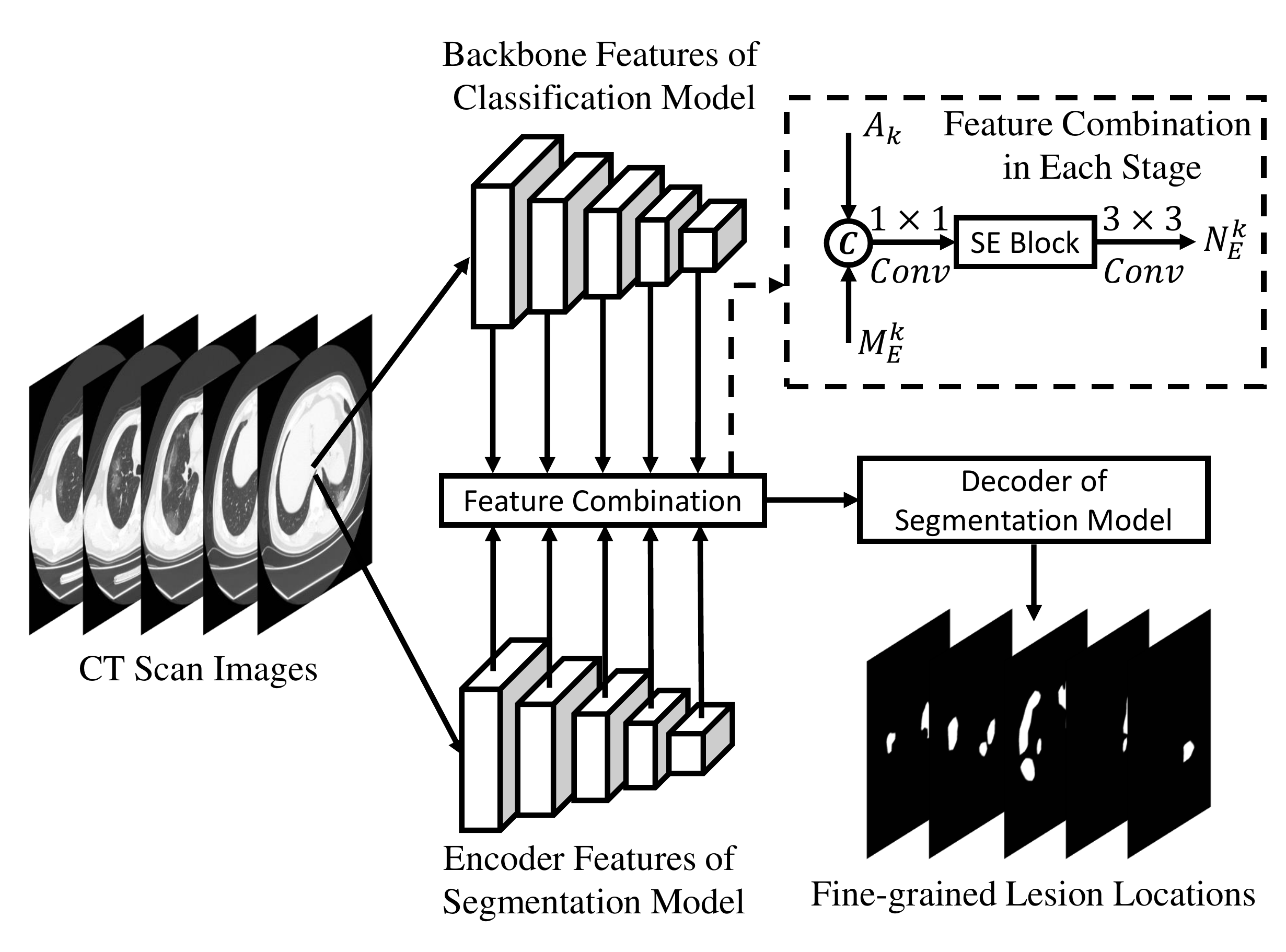}
    \end{overpic}
	\caption{\textbf{Combination of the segmentation and classification models.} We combine the encoder features of the segmentation model with the backbone features of the classification model.
    }
    \label{fig:Combination}
\end{figure}
\vspace{0.01mm}
}

\CheckRmv{
\begin{figure*}[t!]
	\centering
    \small
	\begin{overpic}[width=\textwidth]{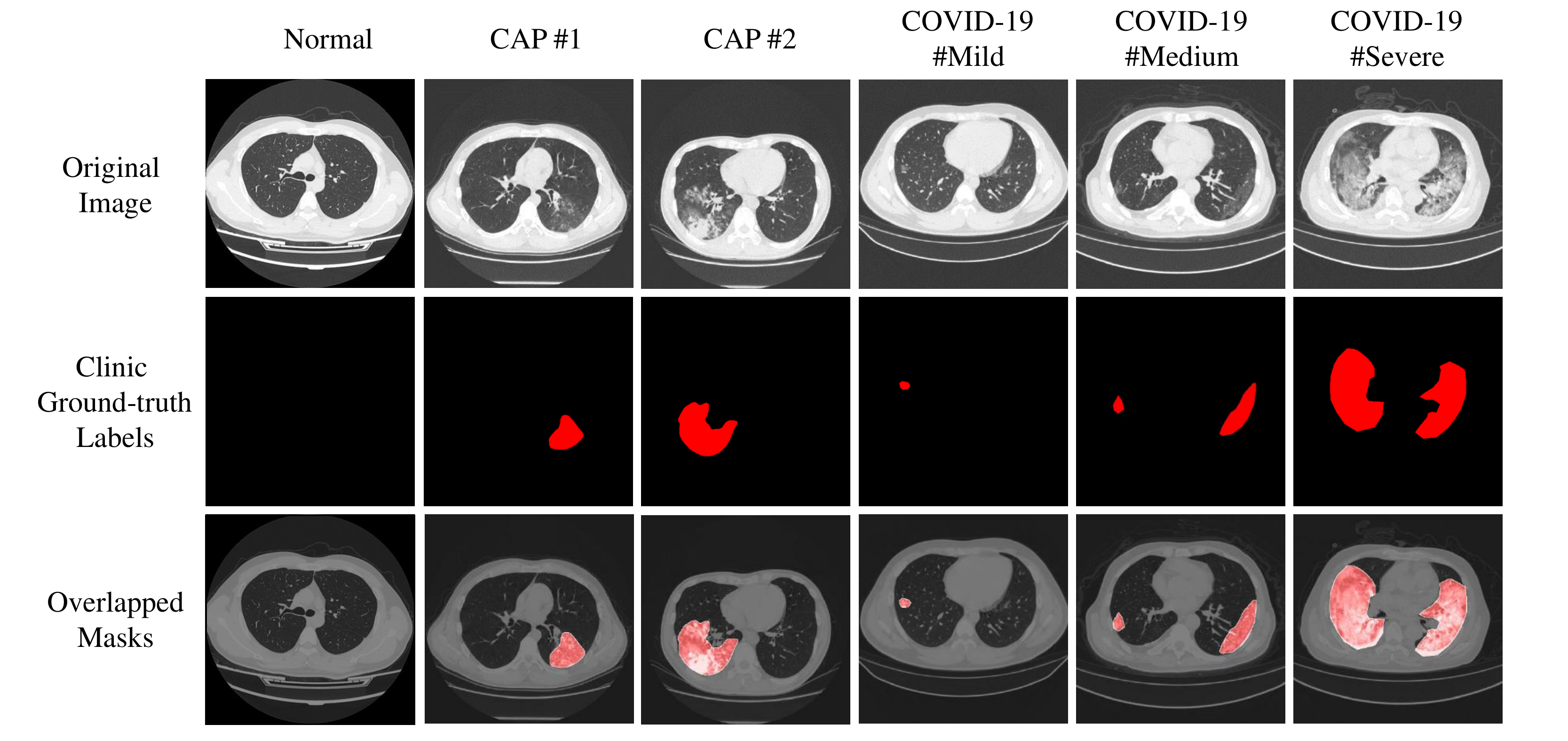}
    \end{overpic}
    \vspace{-10pt}
	\caption{\small \textbf{Examples of our \ourdataset~dataset}, including CT scan images and labels of a normal person ($1^{st}$ column), two community-acquired pneumonia (CAP) cases (2$^{nd}$ and 3$^{rd}$ columns), and three COVID-19 patients from mild to severe ($4^{th}\sim 6^{th}$ columns).}
    \label{fig:Examples}
\end{figure*}
}

\subsubsection{Attentive Feature Fusion}\label{sec:aff}
Traditional fusion strategy of top-down decoders~\cite{ronneberger2015unet, lin2017feature} treats the input feature maps equally.
To better aggregate the feature maps, we propose an Attentive Feature Fusion (AFF) strategy.
In our AFF fusion strategy, the feature map with a smaller size is more valued.
As shown in \figref{fig:TwoModules} (b), the input feature maps $M_E^i$ and $M_D^{i+1}$ in the current stage are reduced to half size via $1\times 1$ convolution layers.
Then the reduced $M_D^{i+1}$ is up-sampled by bilinear interpolation to output a double-sized feature map.
We concatenate the two outputs and apply the SE block (also used in GAM) to produce an enhanced feature map.
This feature map is then concatenated with the feature map of doubly up-sampled output in the previous stage.
After the concatenation, we use another SE block to enhance the feature map again.
After each SE block, we use a $3\times3$ convolution layer, with the same number of channels as the input, as the transition layer.
A $1\times 1$ convolution layer with a single neuron 
will be used to predict one feature map as the side-output of the current stage.

\subsubsection{Combination with the Classification Model}
\label{sec:combine}
As described above, we have designed two models, one for COVID-19 classification and the other one for COVID-19 opacification segmentation.
However, they are separately working on the diagnosis system, and there might be a way to combine them together for better performance.
Inspired by this, we leverage the features of the classification model to enhance the features of the segmentation model. 
As shown in \figref{fig:Combination}, we merge the feature maps of each stage 
from the encoder of the segmentation model and the backbone of the classification model together.
The feature maps of the encoder of the segmentation model are 
$M_E^1, M_E^2, M_E^3, M_E^4, M_E^5$ as defined in \secref{sec:pipeline}.
The Res2Net~\cite{pami20Res2net} backbone of the classification model has five stages 
and we use the last feature maps $A_k$ of stage $k\in [1,5]$ for the feature combination. 
In merging the features of stage $k$, we have two feature maps $A_k, M_E^k$ for the merge.
We first resize the smaller one $A_k$, making it the same size as the larger one $M_E^k$, and concatenate them together. 
Then, we apply a simple $1\times 1$ convolution layer for the feature channel reduction, 
making the output feature maps the same number of channels as $M_E^k$. 
Such $1\times 1$ convolution layer is followed by a SE block with a reduction rate of 4. 
At last we use a $3\times 3$ convolution layer of the same number of input and output channels as the transition layer. 
The output $N_E^k$ will be regarded as the enhanced encoder features and be fed into the decoder of the segmentation model (\figref{fig:seg_pipeline}). 
Then results are predicted as introduced in \secref{sec:pipeline}.

\subsubsection{Deep Supervision Loss}\label{sec:deep_supervision}
Although the final prediction is only from the last side-output, we apply the deep supervision strategy~\cite{lee2015deeply} to all side-outputs with different sizes.
For each side-output, we up-sample it to the size of the ground-truth map, and compute the sum of the standard binary cross-entropy loss and the Dice loss~\cite{milletari2016vnet} as follows:
\begin{equation}
    L = BCE(\mathbf{P},\mathbf{G})  + 1 - \frac {\mathbf{P} \cdot \mathbf{G}}{\|\mathbf{P}\|_{1}+\|\mathbf{G}\|_{1}},
\end{equation}
where the binary cross-entropy (BCE) loss is averaged among all $H\times W$ pixels,
$p_{i,j}$ is the confidence score at pixel $(i,j)$ calculated by a $sigmoid$ function,
and ``$\cdot$'' means the dot product.
$\mathbf{P}$ and $\mathbf{G}$ are predicted map and ground-truth map, respectively, while $\|\mathbf{P}\|_{1}$ and $\|\mathbf{G}\|_{1}$ denote the corresponding $\ell_{1}$ norms.

\CheckRmv{
\begin{figure}[t!]
	\centering
    \small
	\begin{overpic}[width=\columnwidth]{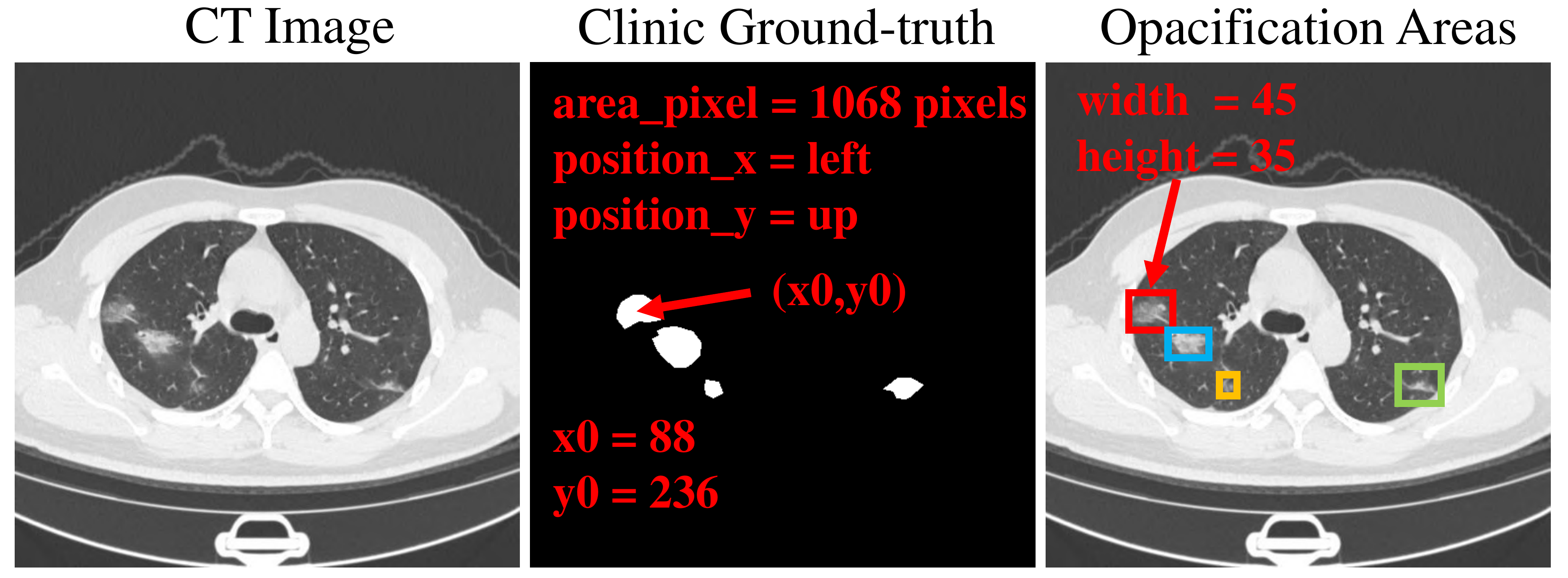}
    \end{overpic}
    \vspace{-4mm}
	\caption{\small \textbf{Illustration of diverse information} about opacification areas (in pixels), location (x0,y0), position (left, up), and width/height of opacification areas in our \ourdataset~dataset.}
    \label{fig:MultipleLabels}
\end{figure}
}

\subsection{Joint Diagnosis}
An explainable classifier or accurate segmentation model itself could not fully implement comprehensive functions for COVID-19 diagnosis.
Comparing to the segmentation model, our classifier is trained with CT images from both COVID-19 infected and uninfected cases,
benefiting from more training data with lower annotation costs.
Although our classifier can provide explainable lesion location of COVID-19 through activation mapping techniques,
it cannot perform accurate and complete lesion segmentation.
To this end, our segmentation model further provides complementary analysis by discovering the complete lesions in the lung and estimate the severity of the COVID-19 patients.
But annotating vast segmentation labels by experienced radiologists is prohibitively expensive.
To integrate their advantages for better application, we develop a diagnosis system for COVID-19 via joint explainable classification and segmentation models.
In practice, our classification model will first predict whether the CT images of a suspected case to be COVID-19 positive or not.
If the prediction is positive, the suspected case is very likely to be infected by COVID-19.
Our segmentation model will then be performed on the CT images for 
in-depth analysis and to discover the whole opacification areas in each CT image.

\section{Our \ourdataset~Dataset}
\label{sec:dataset}
Data plays an essential role in the deep learning-based AI diagnosis systems.
Currently, there are few publicly available COVID-19 datasets with either large scale samples or fine-grained pixel-level labeling.
To fill in this gap, we construct a new COVID-19 Classification and Segmentation (\textbf{\ourdataset})~dataset.
In this section, we present the data collection, professional labeling, and statistics of our dataset.
\figref{fig:Examples} shows some examples of our \ourdataset~dataset (normal case and COVID-19 cases) and examples of CAP patients. 
\figref{fig:MultipleLabels} presents diverse information in the segmentation set of our \ourdataset~dataset.

\CheckRmv{
\begin{figure*}[ht]
	\centering
    \small
	\begin{overpic}[width=\textwidth]{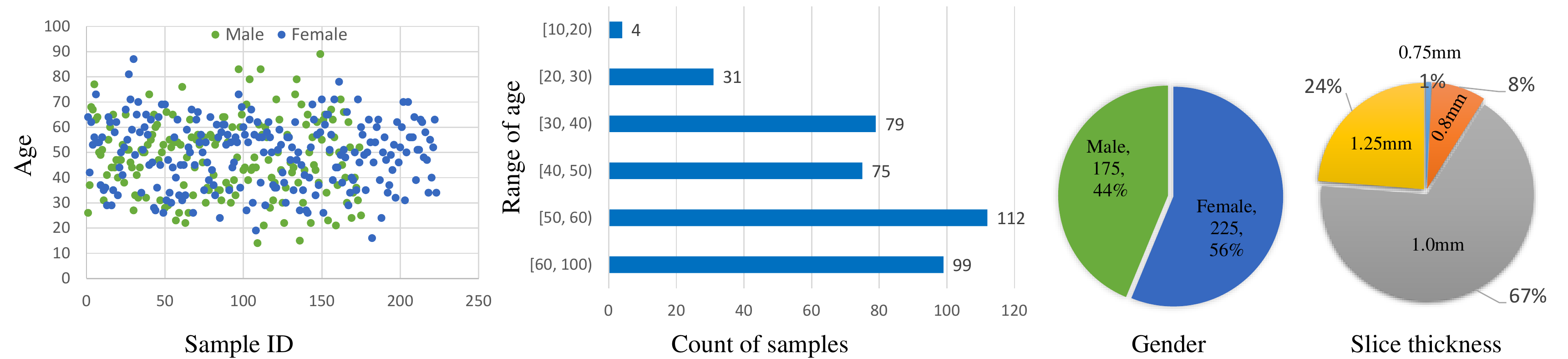}
    \end{overpic}
	\caption{\small \textbf{The age, gender, and slice thickness distribution} of the COVID-19 patients in our \ourdataset~dataset. Zoom in for details.}
    \label{fig:Distribution}
\end{figure*}
}

\CheckRmv{
\begin{figure*}[t!]
	\centering
    \small
	\begin{overpic}[width=\textwidth]{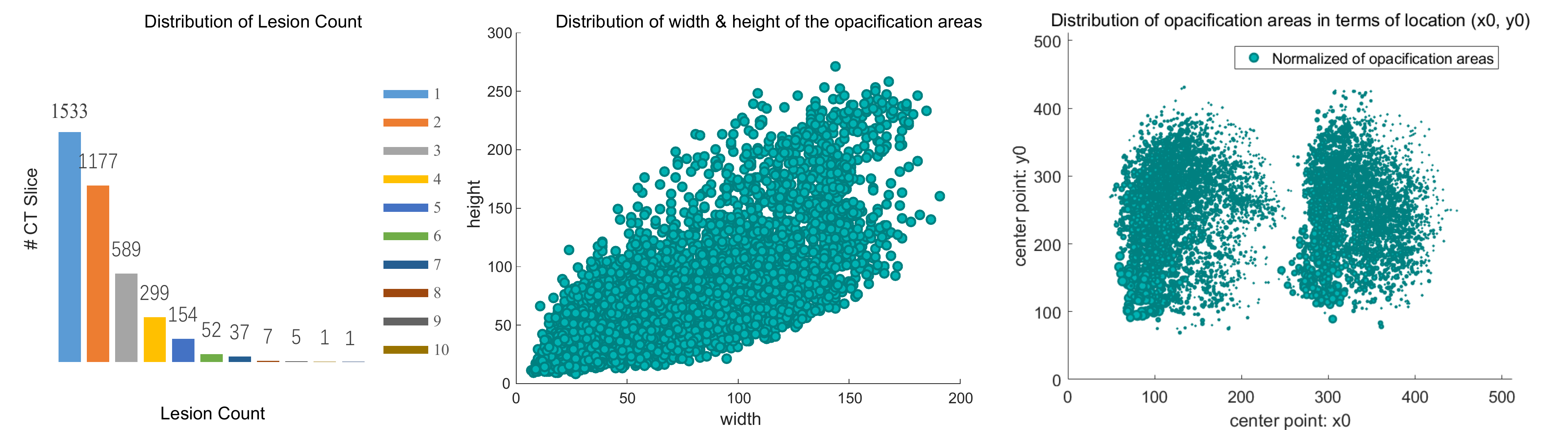}
    \end{overpic}
    \leftline{\hspace{1.in} (a) \hspace{2.1in} (b) \hspace{2.1in} (c)}
	\caption{\small \textbf{Statistics of the segmentation set (200 COVID-19 cases) in our \ourdataset~dataset}. 
	(a) Lesion count distribution. 
	(b) The distribution of width \& height of the opacification areas.
	(c) The relationship between the opacification areas and their locations.}
    \label{fig:Statistics}
\end{figure*}
}

\subsection{Data Collection}
To protect the patients' privacy, we omit their personal information in our dataset construction.
We collected 144,167 CT scan images from 750 cases, 400 of which are positive cases of COVID-19, 
and the other 350 cases are negative, all confirmed by RT-PCR tests.
As previous studies~\cite{shi2020large} did, 
we do not take community-acquired pneumonia (CAP) patients (see \figref{fig:Examples}) into consideration.
Although CAP patients may be diagnosed as COVID-19 positive with our 
proposed diagnosis system since CT images of CAP patients also have similar opacifications, 
the threat of CAP is much less than that of COVID-19. 
And our purpose is to quickly develop an automatic diagnosis system and diagnose suspected cases as soon as possible.
Besides, CAP patients can be simply diagnosed as COVID-19 negative with the help of 
the CAP/COVID-19 classifier~\cite{shi2020large}, RT-PCR test, and the experience of doctors.

All involved patients underwent standard chest CT scans.
Each case has 250 $\sim$ 400 CT images, 
and the number of CT images in each case is only determined 
by the type of the CT scanner and its scan settings.
The CT scanners include BrightSpeed, Discovery CT750 HD, LightSpeed VCT, 
LightSpeed16, Revolution CT from GE Medical Systems, Aquilion ONE from Toshiba,
and uCT 780 from United Imaging Healthcare.
The numbers of cases from different scanners are summarized in \tabref{tab:DataCollectDevice}.
The thickness of reconstructed CT slices ranges from 0.75mm to 1.25mm (\figref{fig:Distribution} for more details).

\CheckRmv{
\begin{table}[t]
  \centering
  \footnotesize
  \renewcommand{\arraystretch}{1.0}
  \setlength\tabcolsep{11.0pt}
  \caption{\small The CT scanners and numbers of positive cases.}
  \label{tab:DataCollectDevice}
  \begin{tabular}{c||cc}
  \hline
   Manufacturer & Product Name & \#Cases \\
  \hline
  \hline
  GE Medical Systems & Revolution CT & 1\\
  GE Medical Systems & LightSpeed VCT & 6\\
  GE Medical Systems & Discovery CT750 HD & 12\\
  GE Medical Systems & BrightSpeed & 12\\
  Toshiba & Aquilion ONE & 33\\
  GE Medical Systems & LightSpeed16 & 64\\
  United Imaging Healthcare & uCT 780 & 272\\
  \hline
  \end{tabular}
\end{table}
}

\subsection{Professional Labeling}
We provide two aspects of labels for the collected CT scan images in our \ourdataset~dataset, so as to implement joint classification and segmentation tasks.
As mentioned above, our dataset is divided into 400 COVID-19 cases and 350 uninfected cases.
For the segmentation task, we perform professional labeling through the following strategies:
\begin{itemize}

\item In order to save their labeling time, the radiologists only select at most 30 discrete CT scan images for each patient, in which the infections are observed for further annotation.
In this step, our goal is to label every opacification area with pixel-level annotations.

\item To generate high-quality annotations, we first invite a  radiologist to mark as many opacification areas as possible based on his/her clinical experience.
Then we invite another senior radiologist to refine the labeling marks several times for cross-validation.
Some inaccurate labels are fixed after this step.    
\end{itemize}

By implementing the above annotation procedures, we finally obtain \PixelAnnoNumberPos pixel-level labeled CT scan images of 200 COVID-19 patients with a resolution of 512$\times$512.
\PatientAnnoNumberPos CT images of the other 200 COVID-19 patients are without pixel-level annotation due to the shortage of radiologists, but such data will be used in classification tests.
As can be seen in the last three columns of \figref{fig:Examples}, our \ourdataset~dataset covers different levels, \ie, mild, medium, and severe, of COVID-19 cases.

\subsection{Dataset Statistics}

\myPara{Age}. The 400 COVID-19 patients (175 males and 225 females) range from 14 to 89 years, with an average age of 48.9 years. 
\figref{fig:Distribution} shows the distribution of ages, the counts of samples in age ranges, and the gender percentages.

\myPara{Lesion count}.
As shown in \figref{fig:Statistics} (a), we illustrate the distribution of lesion counts.
We observe that the lesion count distributes from 1 to 10 in each CT scan image.

\myPara{Opacification areas}.
We plot the widths and heights of the opacification areas in \figref{fig:Statistics} (b).
The ranges of width and height are $7\sim191$ and $8\sim 271$, respectively, showing diverse distributions.

\myPara{Location}. 
We also show the relationship between each opacification area and the corresponding central location (x0, y0) in \figref{fig:Statistics} (c).
As can be seen, the normalized opacification areas range from the smallest size (35/28452 pixels) to the largest size (28452/28452 pixels). 
It also shows that, in our \ourdataset~dataset, the opacification areas are evenly distributed in diverse locations, which are also evenly distributed in the lungs.
\section{Experiments}
\label{sec:experiment}

\newcommand{\addFig}[1]{{\includegraphics[height=.135\textwidth]{am_results/#1.jpg}}}
\newcommand{\addFigs}[1]{}
\CheckRmv{
\begin{figure*}[t]
  \centering
  \small
  \renewcommand{\arraystretch}{0.5}
  \setlength{\tabcolsep}{0.2mm}
  \begin{tabular}{cccccccc}
    \rotatebox[origin=l]{90}{~~(a) AM origin}&
    \addFig{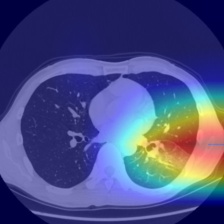}&
     \addFig{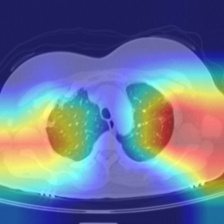}&
     \addFig{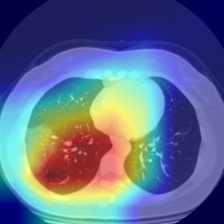}&
    \addFig{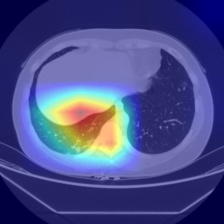}&
     \addFig{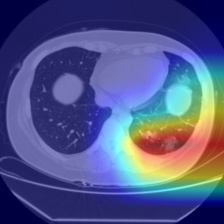}&
    \addFig{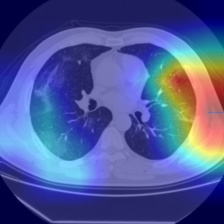}&
    \addFig{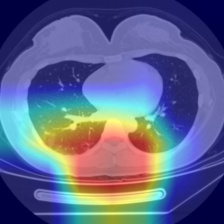}
    \\
   \rotatebox[origin=l]{90}{~(b) AM mixing}&
   \addFig{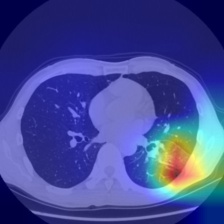}&
    \addFig{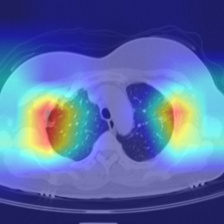}&
    \addFig{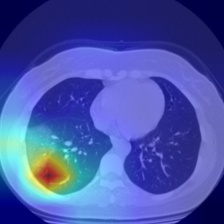}&
   \addFig{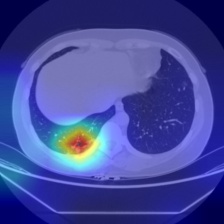}&
    \addFig{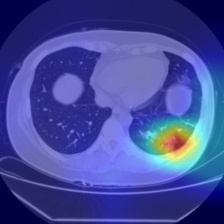}&
   \addFig{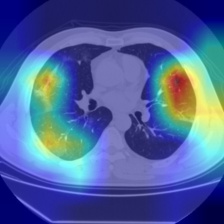}&
   \addFig{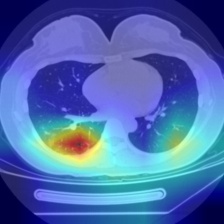} 
  \\
   \rotatebox[origin=l]{90}{~(c) Seg guiding}&
   \addFig{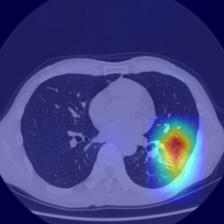}&
    \addFig{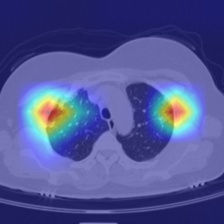}&
    \addFig{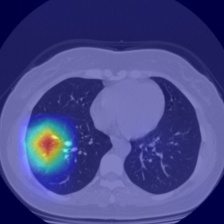}&
   \addFig{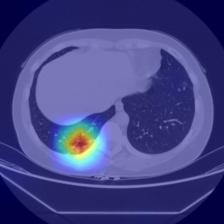}&
    \addFig{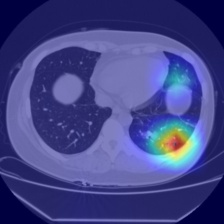}&
   \addFig{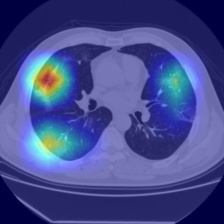}&
   \addFig{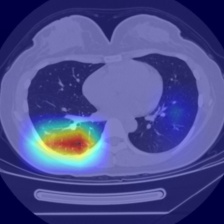} 
   \\
   \rotatebox[origin=l]{90}{~~~(d) Seg. GT}&
   \addFig{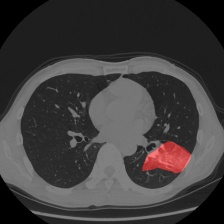}&
    \addFig{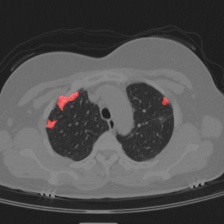}&
    \addFig{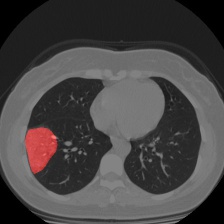}&
   \addFig{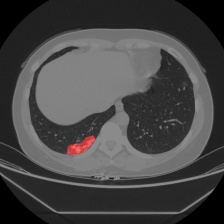}&
    \addFig{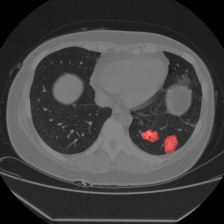}&
   \addFig{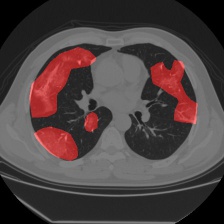}&
   \addFig{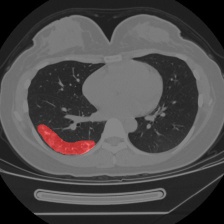}
   \\
  \end{tabular}
  \caption{\textbf{Visualizations of activation mapping (AM)}.
  AM origin (mixing) means the AM of models trained without (with) image mixing technique~\cite{zhang2018mixup}. 
  Seg guiding means the AM of models trained with the segmentation loss $L_{seg}$.}
  \label{fig:am}
\end{figure*}
}

\subsection{Experimental Settings}

\myPara{Training/Test Protocol}.
For the segmentation task, our training set contains 2,794 images from 150 COVID-19 patients and the test set has 1,061 images from the other 50 COVID-19 cases.
For the classification task, the training set contains the 2,794 images from the 150 COVID-19 infected cases in the segmentation set.
In addition, we randomly pick 150 uninfected cases with 7,500 CT images as negative cases for training.
The test set contains the 64,711 images of the other randomly selected 200 infected cases and the 68,041 images from 200 uninfected cases.

\myPara{Evaluation Metrics}.
For the classification task, we adopt the widely used metrics of specificity and sensitivity as suggested by~\cite{TBX11K}.
For the segmentation task, we use two standard metrics, \ie, Dice score~\cite{shan+2020lung} and Intersection over Union (IoU).
To provide a more comprehensive evaluation, we further use 
the widely used metric enhanced alignment measure ($E_\phi$)~\cite{Fan2018Enhanced}.

\myPara{Comparison methods}.
On the classification task, we compare our classification model with or without the image mixing technique~\cite{zhang2018mixup}.
On the segmentation task, to provide an in-depth evaluation of our \ourmodel~model, we compare it with versatile cutting-edge models, \ie, the U-Net~\cite{ronneberger2015unet} for medical imaging and the DSS~\cite{hou2019deeply}, PoolNet~\cite{liu2019simple}, and EGNet~\cite{zhao2019egnet} for saliency detection.

\subsection{Implementation Details}
In our \ourmodel~system, the classification and segmentation models are trained separately.
We implement our system via the PyTorch \cite{paszke2019pytorch} and Jittor \cite{hu2020jittor} framework.
For the classification model, we train it with a batch-size of 256 on 4 GPUs.
The CT images are resized into 224 $\times$ 224 for computational efficiency.
We adopt the SGD optimizer with the initial learning rate of 0.1, 
divided by 10 in every 30 epochs.
The classifier is trained with 100 epochs.
For data augmentation, we use the random~horizontal~flip and random~crop, and the image mixing technique~\cite{zhang2018mixup} to alleviate the data bias. 
The $\alpha$ in the Beta distribution of image mixing is set as 0.5.

For the segmentation model, the number of CT images in each mini-batch is always 4, and the size of the input CT images is unchanged as $512\times 512$.
The backbone of our segmentation model is pretrained on ImageNet\cite{imagenet}.
The atrous rates of four atrous convolutions in two sequential GAMs are $\{1,3,6,9\}$ and $\{1,2,3,4\}$, respectively.
The initial learning rate is $2.5\times 10^{-5}$.
We adopt the $poly$ learning rate policy that the actual learning rate will be multiplied by a factor $(1-\frac{ cur\_iter}{max\_iter})^{power}$, where the power is 0.9.
The segmentation model is trained with 21000 iterations.
We employ the Adam~\cite{kingma2014adam} optimizer and set $\beta_1$, $\beta_2$ as 0.9 and 0.999, respectively.
For data augmentation, we use random~horizontal~flip and random~crop.
When combined with the classification model, the classification model has been pretrained on our classification training set with pixel-level annotations.

\subsection{Results}

\myPara{Activation mapping on explainable classification}.
\figref{fig:am} shows the visualization of activation mapping (AM) of our classification branch trained with or without image mixing~\cite{zhang2018mixup}.
At first, we train our classification model and 
achieve good performance in terms of sensitivity and specificity.
But we find that 
The AM of our classification model initially trained with random horizontal flip and random crop (\figref{fig:am} (a)) not only covers the lesion areas, 
but also presents unrelated areas.
If this problem is not solved, 
an automatic diagnosis system with an overfitted classification network 
is very harmful to clinical diagnosis.
To solve this problem, we investigated and identified that the image mixing technique could solve this problem.
By introducing the image mixing technique~\cite{zhang2018mixup}, 
the AM of our classification model provides more accurate locations 
of the opacification areas as shown in \figref{fig:am} (b).
Moreover, \figref{fig:am} (c) indicates the AM of models trained with the help of pixel-level supervision 
(segmentation loss $L_{seg}$ as introduced in \secref{sec:cls_pixel}).
The AM of models becomes more accurate and specific in locating the opacifications.
However, 
the improvements of adding segmentation loss $L_{seg}$ in classification performance can be ignored,
potentially due to saturated classification accuracy (No.3, \tabref{tab:cls_result}).

\myPara{Performance on explainable classification}.
During the inference, AM assists the medical experts using our \ourmodel~system to judge whether the prediction is correct or not.
For each patient, opacifications can be found in some of its CT images and many images may have no opacifications. 
So we set a threshold for the classification.
When the number of CT images from a suspected patient is larger than a threshold, the patient is diagnosed as COVID-19 positive.
Changing the threshold enables our model to achieve a trade-off between sensitivity and specificity.
\tabref{tab:cls_result} shows the results of the classification model under different thresholds on the test set of our \ourdataset~dataset. 
One can see that our model is very robust to the changing of thresholds, and achieves a sensitivity of 95.0\% and a specificity of 93.0\% when the threshold is 25.
However, AM could not provide accurate segmentation of opacification areas (if any exist).
Subsequently, we further employ our segmentation model to discover the opacification areas in the CT images of COVID-19 patients.

\CheckRmv{
\begin{table}[t]
    \centering
    \small
    \renewcommand{\arraystretch}{1.0}
    \setlength\tabcolsep{15pt}
    \caption{\small Sensitivity and specificity of our classification model under different thresholds.
    We set the threshold as 25 (the gray row) in the final setting. 
    }
    \vspace{-2mm}
    \begin{tabular}{cc||cc}
    \hline
    No. & Threshold & Sensitivity   & Specificity    \\ 
    \hline
    \hline
    1 & 15        & 96.0\%   &  91.5\%     \\
    2 & 20       & 95.0\%   &  92.0\%     \\
    \rowcolor[rgb]{0.9,0.9,0.9}
    3 & \textit{25}        & \textit{95.0\%}   &  \textit{93.0\%}     \\
    4 &30        & 94.5\%   &  93.5\%     \\
    \hline
    \end{tabular}
    \label{tab:cls_result}
\end{table}
}

\CheckRmv{
\begin{table}[t]
    \centering
    \small
    \renewcommand{\tabcolsep}{2.4mm}
    \caption{Ablation Study for the proposed EFM and AFF in the segmentation model. The baseline is the VGG16-based segmentation model without EFM\&AFF (No. 1). 
    We add EFM and AFF separately and show the effectiveness of them (No. 2 and No. 3). The No. 4 result is the complete version of the segmentation model.}
    \label{tab:ablation_module}
    \begin{tabular}{ccc|ccc} \hline
        No. & EFM & AFF & Dice & IoU & $E_\phi$\\ \hline\hline
       1  &  &  & 71.0\% & 57.7\% & 88.0\% \\
       2  & \ding{52} &  & 74.3\% & 61.4\% & 88.9\%  \\
       3  &  & \ding{52} & 75.9\% & 63.4\% & 90.9\% \\
       4  & \ding{52} &  \ding{52} & \textbf{77.5\%} & \textbf{65.4\%} & \textbf{92.0\%} \\
      \hline 
    \end{tabular}
\end{table}
}

\CheckRmv{
\begin{table}[t]
    \centering
    \small
    \renewcommand{\tabcolsep}{2.4mm}
    \caption{Ablation Study for the combination between the segmentation model and the classification model. 
    The baseline segmentation results are generated using the segmentation 
    model only (No.1). 
    After additionally adding features from the classification model,
    we achieve 1.0\% improvement in terms of the Dice metric (No.2).
    }\label{tab:ablation_cls+seg}
    \begin{tabular}{ccc|ccc} \hline
        No. & SEG & +CLS & Dice & IoU & $E_\phi$\\ \hline\hline
       1  & \ding{52} &   & 77.5\% & 65.4\% & 92.0\% \\
       2  & \ding{52} &  \ding{52} & \textbf{78.5\%} & \textbf{66.4\%} & \textbf{92.7\%} \\
      \hline 
    \end{tabular}
\end{table}
}

\CheckRmv{
\begin{table}[t]
  \centering
  \small
  \renewcommand{\arraystretch}{1.0}
  \setlength\tabcolsep{7.0pt}
  \caption{\small Quantitative results on our segmentation test set. 
  }\label{tab:SegResult}
  \begin{tabular}{r|r||c|c|c}
  \hline
   Methods & Publication & Dice & IoU & $E_\phi$\\
  \hline
  \hline
  U-Net~\cite{ronneberger2015unet} & MICCAI'15 & 65.1\% & 54.1\% & 79.7\% \\
  DSS~\cite{hou2019deeply} &  TPAMI'19 & 65.7\%  & 51.7\% & 79.9\%  \\
  EGNet~\cite{zhao2019egnet} & ICCV'19 & 69.3\%  & 55.4\% & 83.6\%  \\
  PoolNet~\cite{liu2019simple} &  CVPR'19 & 69.7\%  & 55.9\%  & 83.9\%  \\
  \hline
  \ourmodel~(Ours)& Submit'20 & \textbf{78.5\%} & \textbf{66.4\%} & \textbf{92.7\%}  \\
  \hline
  \end{tabular}
\end{table}
}

\myPara{Ablation study on our EFM and AFF in the segmentation branch.}
In \secref{sec:segmentation}
we introduced two novel modules named EFM and AFF for the segmentation. 
EFM is designed to enhance the representation power of our encoder in the segmentation branch.
In the feature fusion stage, AFF is applied and the feature map with a smaller size is more valued 
while the traditional fusion strategy treats the input feature maps equally.
The ablation studies for the proposed EFM and AFF are shown in \tabref{tab:ablation_module}.
The No.1 result is the baseline performance without EFM and AFM.
After applying the proposed EFM and AFF separately to the baseline, the performance has 3.3\% and 4.9\% improvement in terms of the Dice metric.
So both EFM and AFF are very helpful for the segmentation branch.
When combining EFM with AFF, we achieve 6.5\% higher results in terms of the Dice metric.
The improvement in terms of the IoU and E-measure~\cite{Fan2018Enhanced} metric is similar to that of the Dice metric.
Hence, the proposed EFM and AFF are very beneficial for the segmentation model.

\newcommand{\AddImg}[1]{\includegraphics[width=.1375\linewidth]{#1}}
\CheckRmv{
\begin{figure*}[t]
\centering
\AddImg{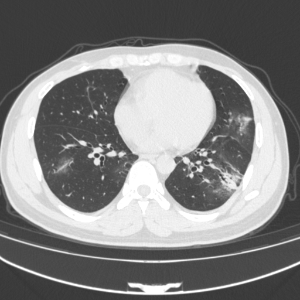} \hfill \AddImg{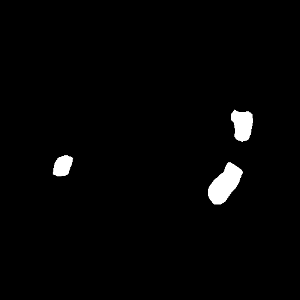} \hfill
\AddImg{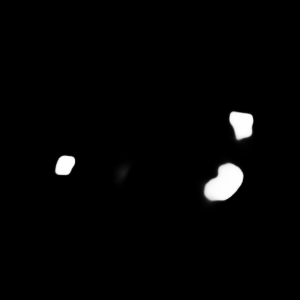} \hfill \AddImg{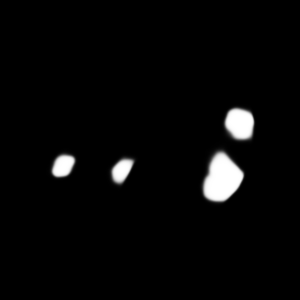} \hfill 
\AddImg{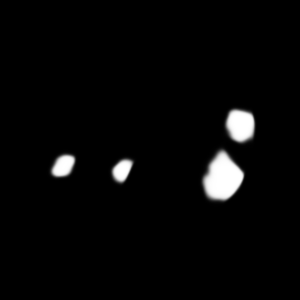} \hfill \AddImg{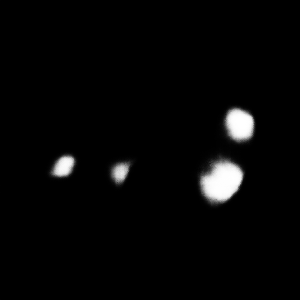} \hfill \AddImg{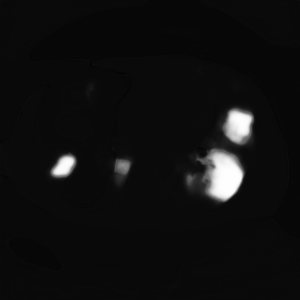}
\\ \vspace{0.03in}
\AddImg{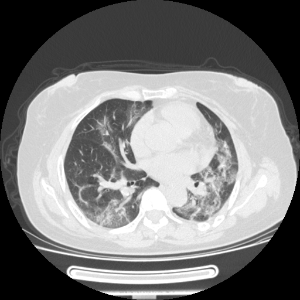} \hfill \AddImg{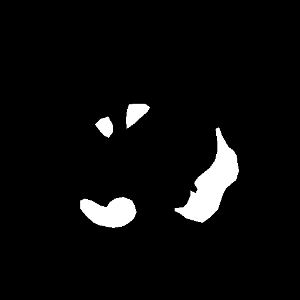} \hfill
\AddImg{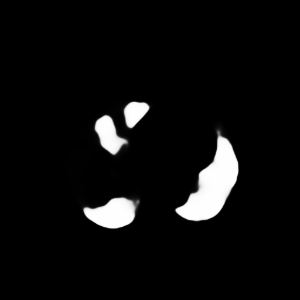} \hfill \AddImg{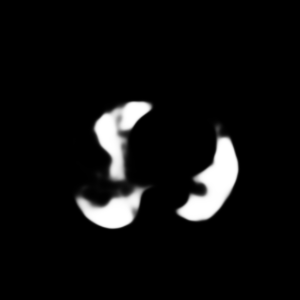} \hfill
\AddImg{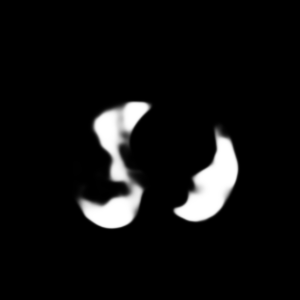} \hfill \AddImg{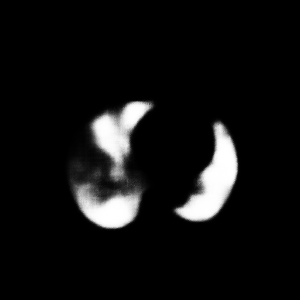} \hfill \AddImg{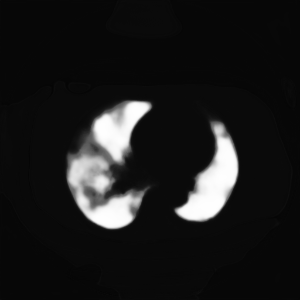}
\\ \vspace{0.03in}
\AddImg{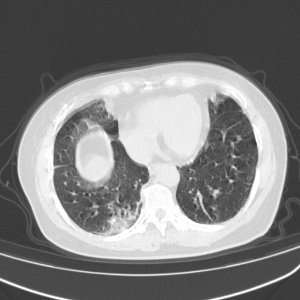} \hfill \AddImg{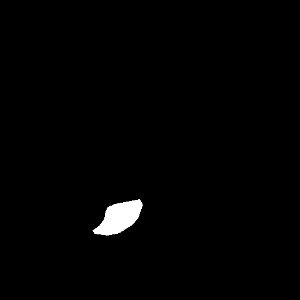} \hfill
\AddImg{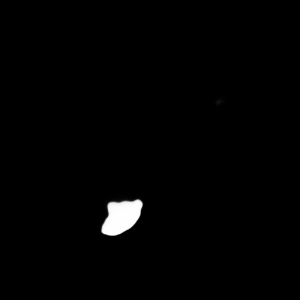} \hfill \AddImg{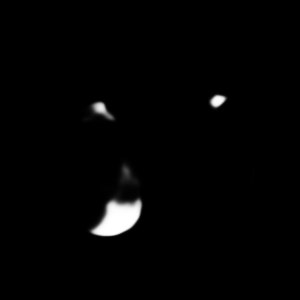} \hfill
\AddImg{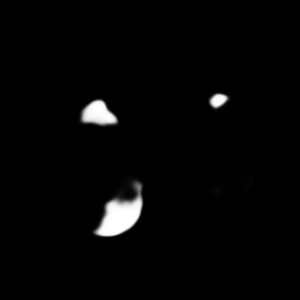} \hfill \AddImg{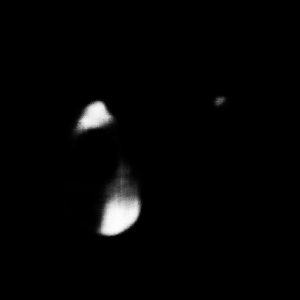} \hfill 
 \AddImg{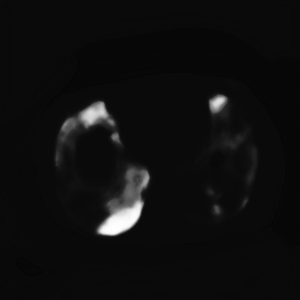}
\\ \vspace{0.03in}
\leftline{\footnotesize \hspace{0.2in} CT image \hspace{0.65in} GT \hspace{0.78in} Ours
\hspace{0.6in} PoolNet\cite{liu2019simple}
\hspace{0.4in} EGNet\cite{zhao2019egnet} \hspace{0.45in} DSS\cite{hou2019deeply}  \hspace{0.5in} U-Net\cite{ronneberger2015unet}}
\caption{\small \textbf{Qualitative comparisons of different methods on our segmentation test set}.
The first, second, and third rows show the comparison results on CT images 
with different lesion areas from the mild, medium, and severe COVID-19 patients, respectively.
}
\label{fig:SegComparison}
\end{figure*}
\vspace{0.01mm}
}

\CheckRmv{
\begin{figure}[t!]
	\centering
    \small
	\begin{overpic}[width=\linewidth]{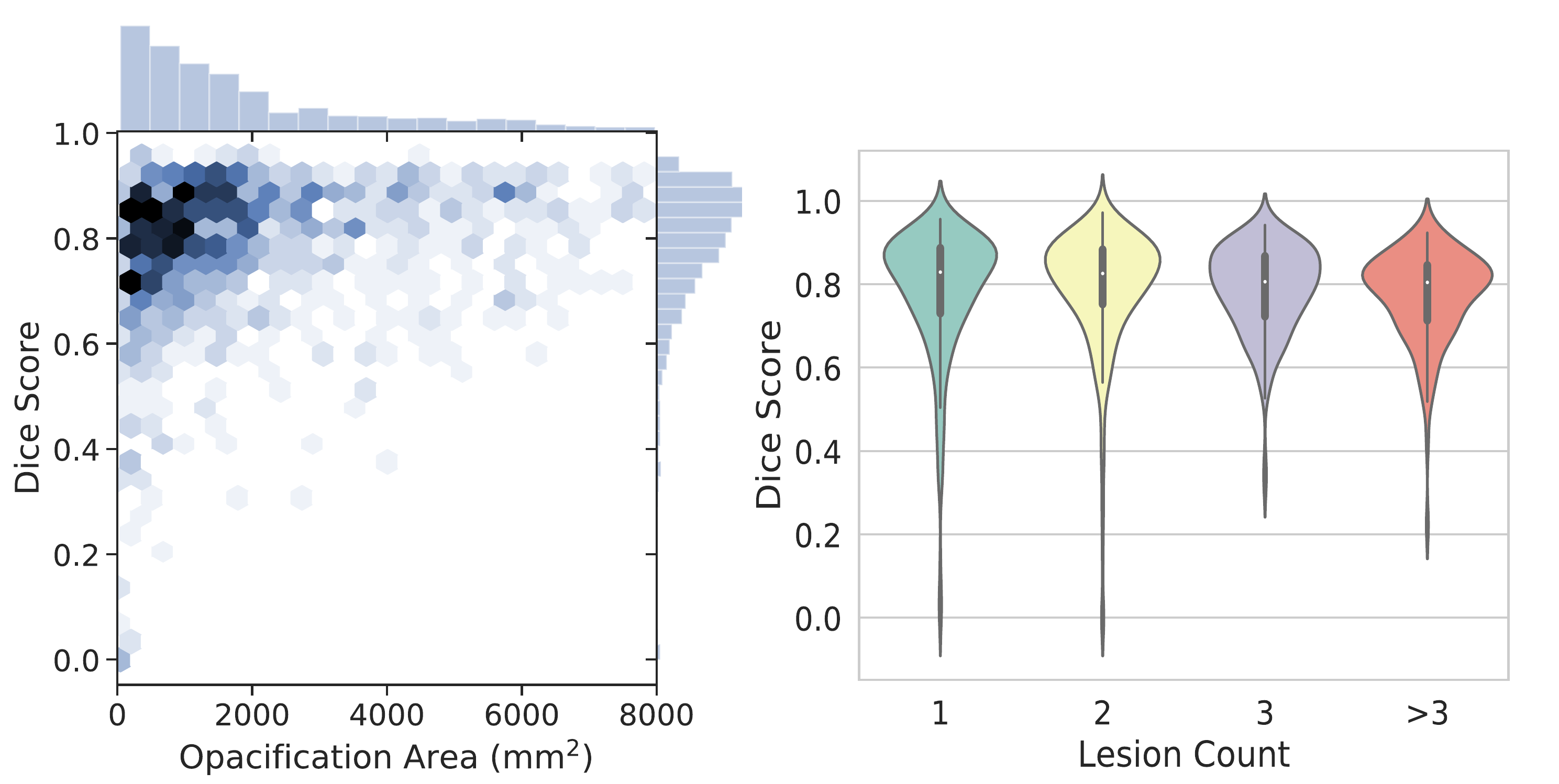}
    \end{overpic}
    \leftline{\hspace{0.7in} (a) \hspace{1.6in} (b)}
	\caption{\textbf{Statistical analysis for our segmentation model on our segmentation test set}.
	(a) The relationship between the opacification area of each CT image and the corresponding Dice score.
	(b) The relationship between the lesion count and the corresponding probability distribution of the Dice score.}
    \label{fig:StaAna}
\end{figure}
\vspace{0.1mm}
}

\myPara{Ablation study on the combination between the segmentation model and classification model.}
As introduced in~\secref{sec:combine},
we combine the classification model with the segmentation model for deriving more abundant features.
To verify such a choice, we run the experiments as shown in \tabref{tab:ablation_cls+seg}.
The baseline is the single segmentation model (No.1, \tabref{tab:ablation_cls+seg}).
But we also observe that the choice of the combination of 
the classification model and segmentation model 
(No.2, \tabref{tab:ablation_cls+seg}) 
has 1.0\% improvement in terms of the Dice metric, 
and shows features of the classification model 
can certainly help the segmentation model predict better results.

\noindent 
\textbf{Comparison of segmentation performance}.
\tabref{tab:SegResult} lists the quantitative comparisons of 4 cutting-edge methods and our model on segmentation.
It can be seen that the proposed model achieves the best result on all three metrics.
It obtains improvements of 8.8\%, 10.5\%, and 8.8\% on Dice score, IoU, and $E_\phi$ over the second-best PoolNet~\cite{liu2019simple}, respectively.
Besides, PoolNet~\cite{liu2019simple} and EGNet~\cite{zhao2019egnet} obtain comparable results on the three metrics. 
U-Net~\cite{ronneberger2015unet} performs better than DSS~\cite{hou2019deeply} in terms of IoU, though they are comparable on the Dice score.
\figref{fig:SegComparison} shows the qualitative results of the comparison methods. 
One can see that the other competitors produce inaccurate or even wrong predictions of the lesion areas in the CT images of mild, medium, and severe COVID-19 infections.
But our segmentation model correctly discovers the whole lesion areas on all levels of COVID-19 infections.

To further study its stability, we perform a statistical analysis of our segmentation model on our segmentation test set.
\figref{fig:StaAna} (a) shows the correlation between the Dice score of our model and the opacification areas of CT images.
Note that the CT images with the opacification area $\ge 8000mm^2$ are not plotted in \figref{fig:StaAna} (a) 
since they only occupy 1.0\% of all CT images in terms of quantity.
We observe that 95.9\% of CT images have the Dice scores in $[0.6,1]$,
while the other 3.3\% of CT images are with Dice scores between $[0.1, 0.6)$ and recognized as bad cases.
Only 0.8\% of CT images suffer from Dice scores of less than 0.1, and they are taken as failure cases.
We also explore the relationship between the lesion count of each slice and the Dice score from a different perspective.
As shown in \figref{fig:StaAna} (b), the probability distribution of the Dice score is little affected by the number of lesion counts in a CT image.
The medium dice score is above 0.8 for 4 different cases of lesion counts, and the 95.0\% confidence interval lies in $[0.5,1]$.
We also observe that the lesion count of failure cases is $\le 2$.
The consistently promising performance on segmenting lesion areas 
and the low probability (0.8\%) of failure confirm the stability of our segmentation model.


\myPara{Diagnosis of time.}
The speed test of the \ourmodel~system is on a single RTX 2080Ti.
Assuming each suspected case has 300 CT images, 
the classification model in \ourmodel~ only costs about 1.0 second to ensure whether infected. If infected, The segmentation model will spend 21.0 seconds on fine-grained lesion segmentation.
Hence, the \ourmodel~system costs 22.0 seconds for each infected case or 1.0 second for each uninfected case.
Note that the complete RT-PCR test and radiologist CT diagnosis cost about 4 hours and 21.5 minutes, 
respectively, no matter the cases are infected or not.

\section{Future Works}

Recently, transformer \cite{vaswani2017attention}, \ie, a very popular operator for NLP, has also achieved great success for computer vision 
since transformer has an excellent ability of modeling global information.
Some of the representatives \cite{wang2021pyramid, liu2021swin, liu2021transformer, wu2021p2t} 
can largely surpass CNN networks with varies of vision tasks such as 
image classification, object detection, and semantic segmentation.
Therefore, we can enhance our diagnosis system via replacing CNN backbones with transformers.
The novel neural architecture search (NAS) \cite{gu2021dots} can automatically 
optimize the detailed architecture of our framework fastly.
At last, there are some novel CNN visualization techniques for providing better CNN explanations \cite{jiang2021layercam}.

\section{Conclusion}
\label{sec:conclusion}
To facilitate the training of strong CNN models for COVID-19 diagnosis, in this paper, we systematically constructed a large scale COVID-19 Classification and Segmentation (\ourdataset) dataset.
We also developed a Joint Classification and Segmentation (\ourmodel) system for COVID-19 diagnosis.
In our system, the classification model identified whether the suspected patient is COVID-19 positive or not, along with convincing visual explanations.
It obtained a 95.0\% sensitivity and 93.0\% specificity on the classification test set of our \ourdataset~dataset.
To provide complementary pixel-level prediction, we implemented a segmentation model to discover
fine-grained lesion areas in the CT images of COVID-19 patients.
Comparing to the competing methods, e.g., PoolNet~\cite{liu2019simple}, our segmentation model achieved an improvement of 8.8\% on the Dice metric.
Our \ourmodel~system is also very stable.
On our segmentation test set, it failed only on 0.8\% images and obtained Dice scores between $[0.6,1]$ for 95.9\% of images.
The online demo of our \ourmodel~diagnosis system for COVID-19 will be available soon.

\section*{Acknowledgment}
This research was supported by Major Project for New Generation of AI 
under Grant No. 2018AAA0100400, NSFC (61922046), 
and Tianjin Natural Science Foundation (17JCJQJC43700, 18ZXZNGX00110).

{
\small
\bibliographystyle{IEEEtran}
\bibliography{covid}
}

\newcommand{\AddPhoto}[1]{\includegraphics[width=1in,keepaspectratio]{#1}}

\begin{IEEEbiography}[\AddPhoto{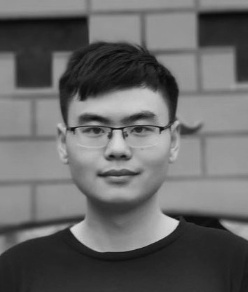}]{Yu-Huan Wu}
is currently a Ph.D. candidate with College of Computer Science 
at Nankai University, supervised by Prof. Ming-Ming Cheng. 
He received his bachelor's degree from Xidian University in 2018. 
His research interests include computer vision
and machine learning.
\end{IEEEbiography}

\begin{IEEEbiography}[\AddPhoto{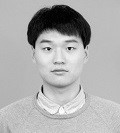}]{Shang-Hua Gao}
 is a Ph.D. student in Media Computing Lab at Nankai University. He is supervised via Prof. Ming-Ming Cheng. His research interests include computer vision, machine learning, and radio vortex wireless communications.
\end{IEEEbiography}

\begin{IEEEbiography}[\AddPhoto{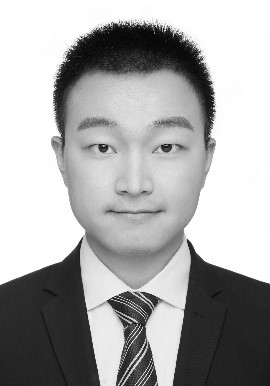}]{Jie Mei}
is a Ph.D. student in College of Computer Science, Nankai University, Tianjin, China.
His research interests include computer vision,
machine learning, and remote sensing image processing.\end{IEEEbiography}

\begin{IEEEbiography}[\AddPhoto{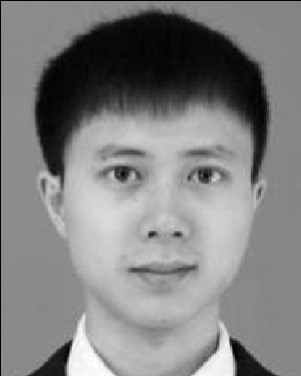}]{Jun Xu}
received his B.Sc. and M.Sc. degrees from School of Mathematics Science, Nankai University, Tianjin, China, in 2011 and 2014, respectively, and the Ph.D. degree from the Department of Computing, Hong Kong Polytechnic University, in 2018. He worked as a Research Scientist at IIAI, Abu Dhabi, UAE. He is currently a Lecturer with School of Statistics and Data Science, Nankai University. More information can be found at \url{https://csjunxu.github.io/}.
\end{IEEEbiography}

\begin{IEEEbiography}[\AddPhoto{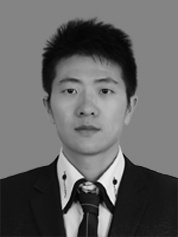}]{Deng-Ping Fan}
 received his Ph.D. degree from the Nankai University in 2019.
He joined Inception Institute of Artificial Intelligence (IIAI) in 2019.
He has published about 20 top journal and conference papers such as
CVPR, ICCV, etc. 
His research interests include computer vision, deep learning, and 
saliency detection, especially on co-salient object detection, 
RGB salient object detection, 
RGB-D salient object detection, and video salient object detection.
\end{IEEEbiography}

\begin{IEEEbiography}[\AddPhoto{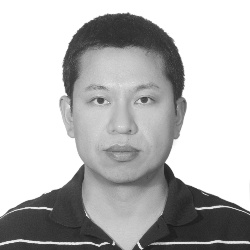}]{Rong-Guo Zhang}
 received his Ph.D. degree majoring in pattern recognition and intelligent systems from Institute of Automation,
 Chinese Academy of Sciences in 2012.
 Now he serves as head of Institute of Advanced Research, 
 Beijing Infervision Technology Co Ltd. 
 His research interests include computer vision, 
 deep learning, and medical image processing.
\end{IEEEbiography}

\begin{IEEEbiography}[\AddPhoto{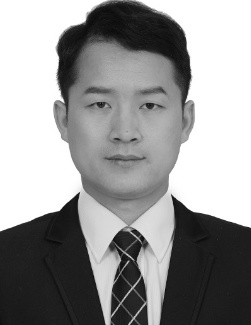}]{Ming-Ming Cheng}
received his Ph.D. degree from Tsinghua University in 2012.
Then he did two years research fellow with Prof. Philip Torr
in Oxford.
He is now a professor at Nankai University, leading the
Media Computing Lab.
His research interests include computer graphics, computer
vision, and image processing.
He received research awards, including ACM China Rising Star Award,
IBM Global SUR Award, and CCF-Intel Young Faculty Researcher Program.
He is on the editorial boards of IEEE TIP.
\end{IEEEbiography}

\vfill
\end{document}